\documentclass[reprint,showpacs,preprintnumbers]{revtex4-1}
\usepackage{graphicx}
\usepackage{longtable}
\usepackage{footnote}
\begin{document}

%\preprint{APS/123-QED}

\title {First-principles study of the optoelectronic properties and photovoltaic absorber layer efficiency of Cu-based chalcogenides}
\author{N. Sarmadian}
%\author{N. Sarmadian$^{a}$}
%\footnotetext{\textit{$^{a}$nasrin.sarmadian@uantwerpen.be}}
\email{nasrin.sarmadian@uantwerpen.be}
\author{R. Saniz}

\author{B. Partoens}

\author{D. Lamoen}

\affiliation{EMAT \& CMT groups, Departement Fysica, Universiteit Antwerpen, Antwerpen, Belgium\\}
%EMAT \& CMT groups, Departement Fysica, Universiteit Antwerpen, Antwerpen, Belgium\\

\begin{abstract}
Cu-based chalcogenides are promising materials for thin-film solar cells with more than 20\% measured cell efficiency. Using first-principles calculations based on density functional theory, the optoelectronic properties of a group of Cu-based chalcogenides Cu$_2$-II-IV-VI$_4$ is studied. They are then screened with the aim of identifying
potential absorber materials for photovoltaic applications. The spectroscopic limited maximum efficiency (SLME) introduced by Yu and Zunger is used as a metric for the screening.
After constructing the current-−voltage curve, the maximum spectroscopy--dependent
power conversion efficiency is calculated from the maximum power output.
The role of the nature of the band gap, direct or indirect, and also of the absorptivity
of the studied materials on the maximum theoretical power conversion efficiency is studied.
Our results show that Cu$_2$II-GeSe$_4$ with II=Cd and Hg, and Cu$_2$-II-SnS$_4$ with II=Cd and Zn have a higher theoretical efficiency compared to the materials currently used as absorber layer.
\end{abstract}

\maketitle

\section{\label{V_introduction}Introduction\protect\\}

The potential applications of the multinary chalcogenide semiconductors in optoelectronics give rise to an intense interest in their design and synthesis that dates back to the
1950s \cite{3J,4J,Cu2ZnGeTe4_exp,127}.
Ternary I-III-VI$_2$ compounds can be generated from binary II-VI chalcogenides through substituting
group II atoms by pairs of group I and III atoms.
Because of the increased chemical and structural flexibility in ternary compounds,
they exhibit a larger variety of optoelectronic properties compared to binary ones \cite{pv6}.
For example, CuGaSe$_2$ has a band gap of 1.68 eV which is lower than that of ZnSe (2.82 eV) \cite{12L}.
This is one of the characteristics of CuGaSe$_2$ that makes it
convenient for application in thin-film solar cells.
Further flexibility is obtained by introducing quaternary chalcogenides and this allows to
engineer the functional properties to satisfy a certain application, e.g. high-efficiency
photovoltaic absorber layers or light emitting diodes.\

\indent
There are two approaches to substitute the cations in ternary I-III-VI$_2$ to design quaternary compounds.
One is to replace two III atoms by one II and one IV atom, forming a I$_2$-II-IV-VI$_4$ compound.
The other one is to replace one I atom and one III atom by two II atoms, forming
II$_2$-I-III-VI$_4$ compounds. Such quaternary chalcogenides with I = $\{$Cu, Ag$\}$, II = $\{$Zn, Cd$\}$,
III = $\{$Ga, In$\}$, IV = $\{$Ge, Sn$\}$, and VI = $\{$S, Se, Te$\}$
have been synthesized by different groups \cite{4J,15L,Cu2ZnGeTe4_exp}.
In particular, Cu-based chalcogenides Cu$_2$-II-IV-VI$_4$ can be found at
the center of various technological innovations.
Among these compounds, Cu$_2$ZnSnS$_4$ (CZTS) and Cu$_2$ZnSnSe$_4$ (CZTSe)
combine promising characteristics for optoelectronic applications
(e.g. direct band gap of 1.0-1.4 eV, a high optical absorption coefficient up to $10^{5} cm^{-1}$,
and a relatively high abundance of the elements \cite{2F,7G,127}).
Such characteristics make them the low-cost alternative to the conventional
photovoltaic materials like Si, CdTe and CuIn$_{1-x}$Ga$_x$Se$_2$ \cite{2F,7G}.
While significant attention has been paid to CZTS and CZTSe \cite{3J,pv5,22L},
most of the other compounds in this family remain relatively unexplored.
Limited theoretical attention has been paid to these chalcogenides,
so their electronic structure and optical properties remain unclear,
which limits their usage in semiconductor devices.
A deeper knowledge of their optoelectronic properties
might bring further improvements in their applications \cite{5H}.\\
\indent
On the one hand, extensive measurements have been performed to study the change of the power
conversion efficiency of the photovoltaic solar cells with respect to the characteristics of
the absorber layers. On the other hand, the conversion efficiency of the solar cells is
investigated theoretically, but very few of such studies calculate the efficiency of the
solar cells using first-principles methods.
Some successful first-principles studies have identified new materials with high conversion efficiency
for PV applications \cite{zunger, perovskite, SLME1}.
Yu and Zunger introduced the "spectroscopic limited maximum efficiency (SLME)"
which is a theoretical power efficiency that can be investigated using first-principles
calculated quantities. They used the SLME parameter as a selection metric to
identify new absorber materials \cite{zunger}.\\

One of the features of the SLME is including the effect of the thickness in the efficiency of the absorber layer which is not taken into account in the well known maximum theoretical efficiency, Shockley–-Queisser (SQ) limit \cite{SQ}. The thickness of the absorber layers in the existing thin film solar cells is few
micrometers \cite{micron_CIGS}. If this thickness could be reduced with only minor loss in
performance of the solar cell, the production costs could be lowered.
Calculating the SLME of a material provides insight about how thin that material can be made with no
significant loss in its efficiency.\\
\indent
One of the main goals of the present manuscript is to investigate how the optoelectronic properties of
the Cu$_2$-II-IV-VI$_4$ compounds change by modifying the material composition.
Moreover, we propose some potential new absorber materials using the SLME parameter.

In the Section~\ref{HT_method} we present the methodology used. Section~\ref{HT_results}
presents our results together with a discussion. We end this work with the Section~\ref{HT_summary},
where we summarize our main findings.

\section{Structure of the chalcogenides materials}\label{structure}

The distribution of the cations within the unit cell of Cu$_2$-II-IV-VI$_4$ depends on
the nature of the group II and IV atoms. For example the kesterite structure is
the most stable phase for Cu$_2$ZnSnSe$_4$ \cite{13F}, while it is reported that
Cu$_2$CdSnSe$_4$ and Cu$_2$HgSnSe$_4$ prefer the stannite structure as the energetically
favorable one \cite{130,145}. Because of the limited number of studies on the stannite structure
of the quaternary Cu-based chalcogenides, the optoelectronic properties of the compounds
are not not well-known. We investigate a group of Cu-based materials that can be found in
the stannite structure: Cu$_2$-II-IV-VI$_4$ with II = $\{$Cd, Hg, and  Zn$\}$, IV = $\{$Sn, and  Ge$\}$,
and VI = $\{$S, Se, and  Te$\}$.\

\noindent
Stannite Cu$_2$-II-IV-VI$_4$ compounds are quaternary complexes with a crystal structure similar to the
zinc-blende structure of ZnS and the kesterite structure of CuInS$_2$.
The stannite primitive cell (space group Ia$\bar4$2m, No. 121) contains 8 atoms.\

\begin{figure}[!htp]
\begin{center}
\includegraphics[width=0.60\hsize]{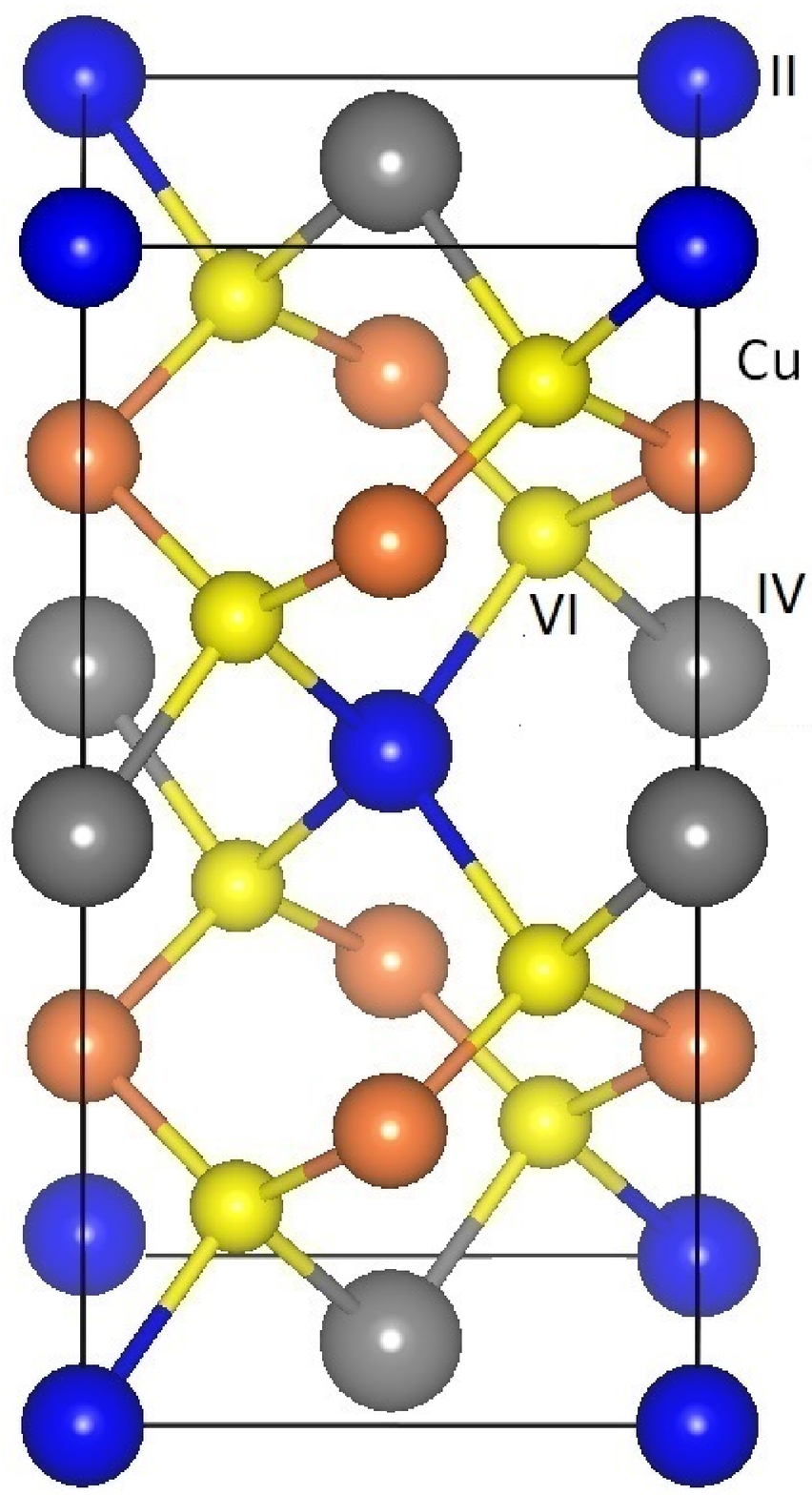}
\caption{\label{fig1} Crystal structure of Cu$_2$-II-IV-VI$_4$-stannite (space group I$\bar4$2m, No. 121).}
\vspace{-0.5cm}
\end{center}
\end{figure}

\noindent
Figure~\ref{fig1} presents the stannite structure. There are alternating cation layers of
mixed II and IV atoms, which are separated from each other by layers of Cu monovalent cations.
Each anions  is tetrahedrally coordinated by four cations. Two equivalent Cu atoms occupy
the 4d Wyckoff position (site symmetry S4), one II atom on 2a, one IV atom on 2b (both II and IV with
D$_{2d}$ symmetry) and four VI atoms on 8i position (site symmetry C$_s$).
In this structure, each anion has thereby three inequivalent bonds(VI-Met) with the cations Met = $\{$Cu, Zn, and Sn$\}$.

%figure1*

\section{Computational Method}\label{HT_method}

Our {\it ab initio} computations are based on DFT \cite{Hohenberg_Kohn1,kohn_sham_1965},
and are carried out using the VASP code \cite{c3,10b,vasp,ch2_78}. We use PAW \cite{c6,PAW}
potentials to describe the electron-ion interaction. We use the generalized gradient approximation (GGA)
to the exchange-correlation potential, in the PBE parametrization \cite{PBE_GGA}
to perform all structural calculations. Electronic structure and optical properties are
calculated using the HSE06 hybrid functional \cite{19}.
An energy cutoff of 350 eV is used for the plane-wave basis set. For structural
relaxation and total energy calculations the Brillouin Zone was sampled using
a 4$\times$4$\times$4 Monkhorst-Pack (MP) grid \cite{MP}. Atomic relaxations were made until residual forces
on the atoms were less than 0.01 eV/{\AA} and total energies were converged to within 1 meV.

In order to calculate the absorption spectra, the imaginary part of
the dielectric function ($\varepsilon(\omega)$) tensor is obtained using the random phase
approximation, as implemented in the VASP code \cite{c9}.
The dielectric function tensor of the studied compounds that have the tetragonal structure
can be described completely by two non-zero independent components,
namely $\varepsilon_{\bot}=\varepsilon_{xx}$, and $\varepsilon_{\|}=\varepsilon_{zz}$ which
corresponds to the dielectric function along the x-, and z-direction, respectively.
The real part of the dielectric function is obtained from the imaginary part through the
Kramers-Kronig relation. We found that it is enough to sample the Brillouin zone using
a 10$\times$10$\times$10 MP grid to obtain a converged $\varepsilon(\omega)$ tensor. The number
of unoccupied bands used here is 3 times the number of occupied bands.

Since the photovoltaic conversion efficiency strongly depends on the band gap,
it is important to get an accurate value from our first-principles calculations.
It is known that standard DFT calculations, using local or semi-local exchange-correlation functionals
such as the local density approximation (LDA) or GGA, seriously underestimate the band gap of
semiconductors \cite{18,19}, while the hybrid functional HSE06 has proven to be capable of giving
close-to-experiment predictions for a large range of compounds including Cu-based compounds \cite{20, zunger}.
Moreover, for a series of compounds, HSE06 provides a dielectric function in much better agreement
with experiment than GGA or LDA functionals \cite{dielectric_exp_PBE_HSE1,dielectric_exp_PBE_HSE2}.

We calculate the power conversion efficiency $\eta$ of an absorber layer which is defined as $\eta=P_m/P_{in}$, where $P_m$ is
the maximum output power density and $P_{in}$ is the total incident
solar power density. $P_m$ can be obtained by numerically
maximizing the $J \times V$ where $J$ is current density and $V$ stands for voltage.
The total current density for a solar cell illuminated under the photon flux $I_{sun}$ at
temperature $T$ is given by $J = J_{sc}-J_{loss}$ \cite{efficiency1}. In this work, the standard AM1.5G
solar spectrum at 25$^{\circ}C$ is used \cite{AM1.5}. Consistently, all parameters that contribute to the SLME and SQ limit are calculated for $T=25^{\circ}C$.

\noindent
The first term in the formula $J = J_{sc}-J_{loss}$ is the short-circuit current density $J_{sc}$ given by
\begin{equation}\label{eq1}
J_{sc}=e\int_0^{\infty}a(E)I_{sun}(E)dE
\end{equation}

\noindent
where $e$ is the elementary charge and $a(E)$ is the
photon absorptivity. The second term is the reverse saturation
current density or loss current density is calculated using the formula

\begin{equation}\label{{eq2}}
J_{loss} = J_0(1-e^{eV/k_BT})
\end{equation}

\noindent
where $k_B$ is the Boltzmann constant, and $T$ is the temperature. $J_0 = J_0^{nr}+J_0^r = J_0^r/f_r$,
corresponding to the total electron-hole recombination current density at equilibrium in the dark.
This recombination includes both nonradiative $J_0^{nr}$ and radiative $J_0^{r}$ current densities,
where $f_{r}$ is the fraction of the radiative recombination current. $f_{r}$ is
approximated by $e^{-\Delta/k_BT}$ where $\Delta = E^{optical}_g - E^{fundamental}_g$ \cite{zunger}.
In equilibrium the rates of emission and absorption through cell surfaces should be equal in the dark.
Thus, the rate of black-body photon absorption from the surrounding thermal bath through the front surface
of the cell gives $J_0^{r}$

\begin{equation}\label{eq3}
J_0^{r} = e\pi\int_0^{\infty}a(E)I_{bb}(E,T)dE
\end{equation}

\noindent
where $I_{bb}$ is the black-body spectrum \cite{efficiency2}. With $a(E)$ modeled as $1-e^{2\alpha(E)L}$,
$\eta$ can be calculated. $\alpha(E)$ is the absorption spectrum of the material and $L$ is the
thin film thickness.\

\indent
We also address some characteristics of the solar cell such as the fill factor (FF).
FF represents the fraction of the maximum power that can be obtained from the cell and
it is the ratio between $P_m$ and $P_{nominal}$. $P_{nominal}$ is simply the product of the highest value
of the voltage and current density of the solar cell, $P_{nominal} = J_{sc} \times V_{oc}$.
The open circuit voltage ($V_{oc}$) which is the voltage of the solar cell under $J = 0$ is calculated
using the formula

\begin{equation}\label{eq4}
V_{oc}=\frac{k_{B}T_{c}}{e} \ln(1+\frac{J_{sc}}{J_0}).
\end{equation}

\section{Results}\label{HT_results}
\subsection{Electronic structure and optical properties}\label{optoelectronic}

In order to calculate the current density, the absorptivity of the compound should be calculated (cf. Eqs.~\ref{eq1} and ~\ref{eq3}). Using the dielectric function, the absorption spectra and absorptivity are calculated.
Figures~\ref{fig-S1}-~\ref{fig-S3} in Appendix \ref{appendix} show the imaginary ($\varepsilon_i$) and real ($\varepsilon_r$) parts of the dielectric function for Cu-based chalcogenides along the x-, and z-direction,
$\varepsilon_\bot$, and $\varepsilon_\|$, respectively. We have noticed that the intensity of the peaks
in $\varepsilon_i$ and the value of the optical dielectric
constant ($\varepsilon_\infty$) along the z-direction are larger than those corresponding to the x-direction.
This results in a higher value of the refractive index along the z-direction compared with the one
along the x-direction. It means that the birefringent studied compounds are optically uniaxial and
they all have positive birefringence.

%figure2*
\begin{figure}[!htp]
\begin{center}
\includegraphics[width=0.49\hsize]{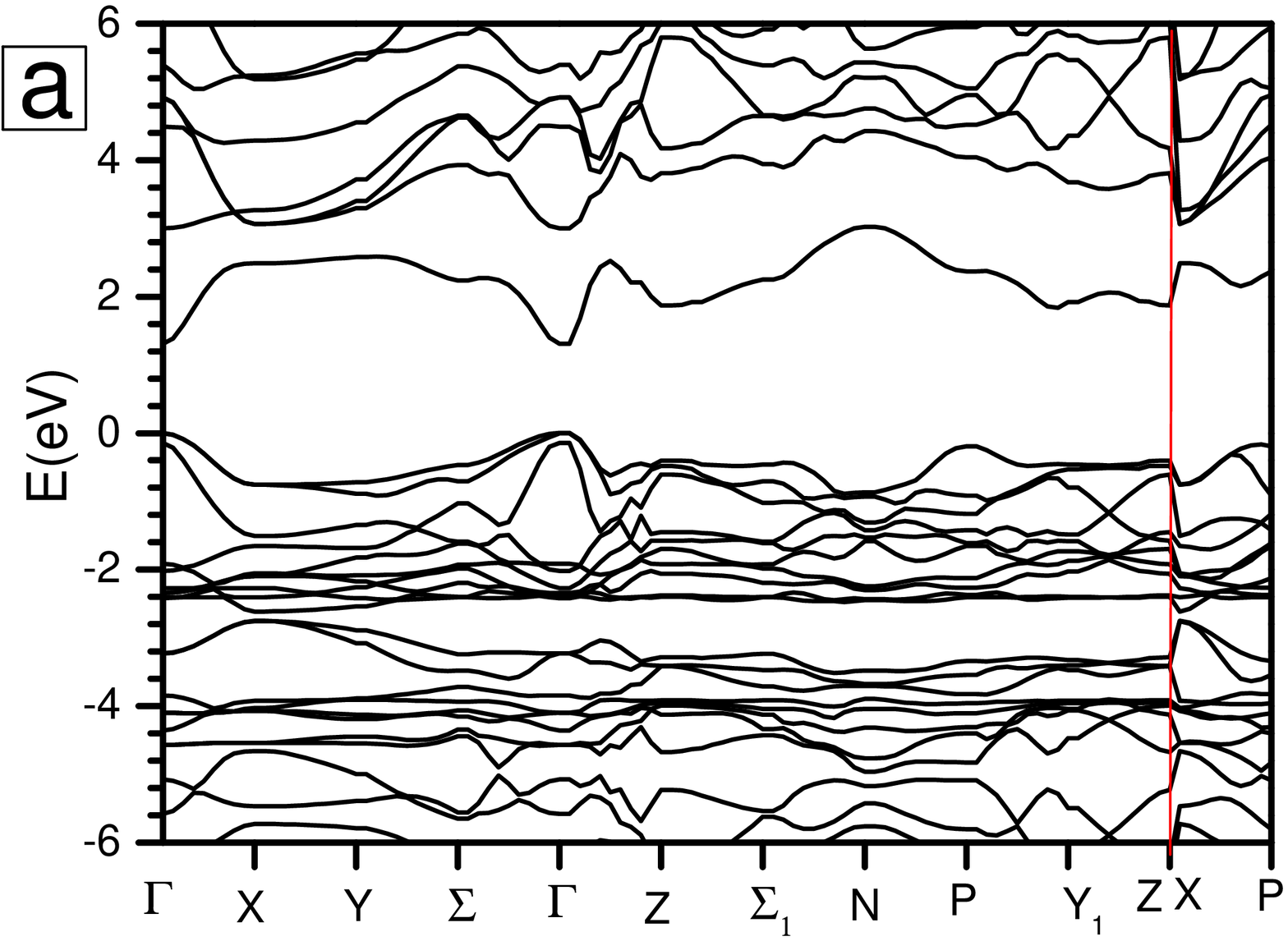}
\includegraphics[width=0.49\hsize]{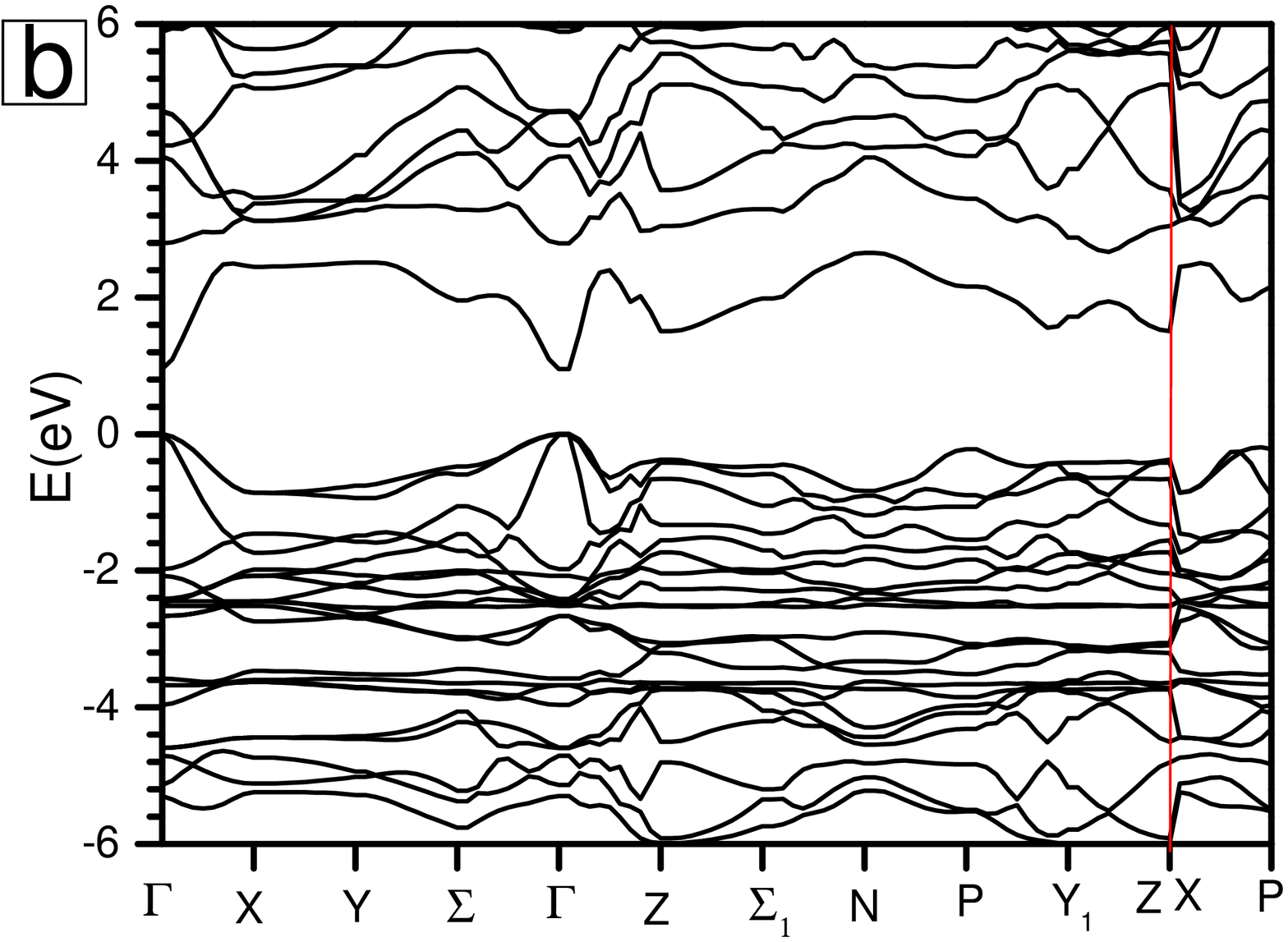}
\includegraphics[width=0.49\hsize]{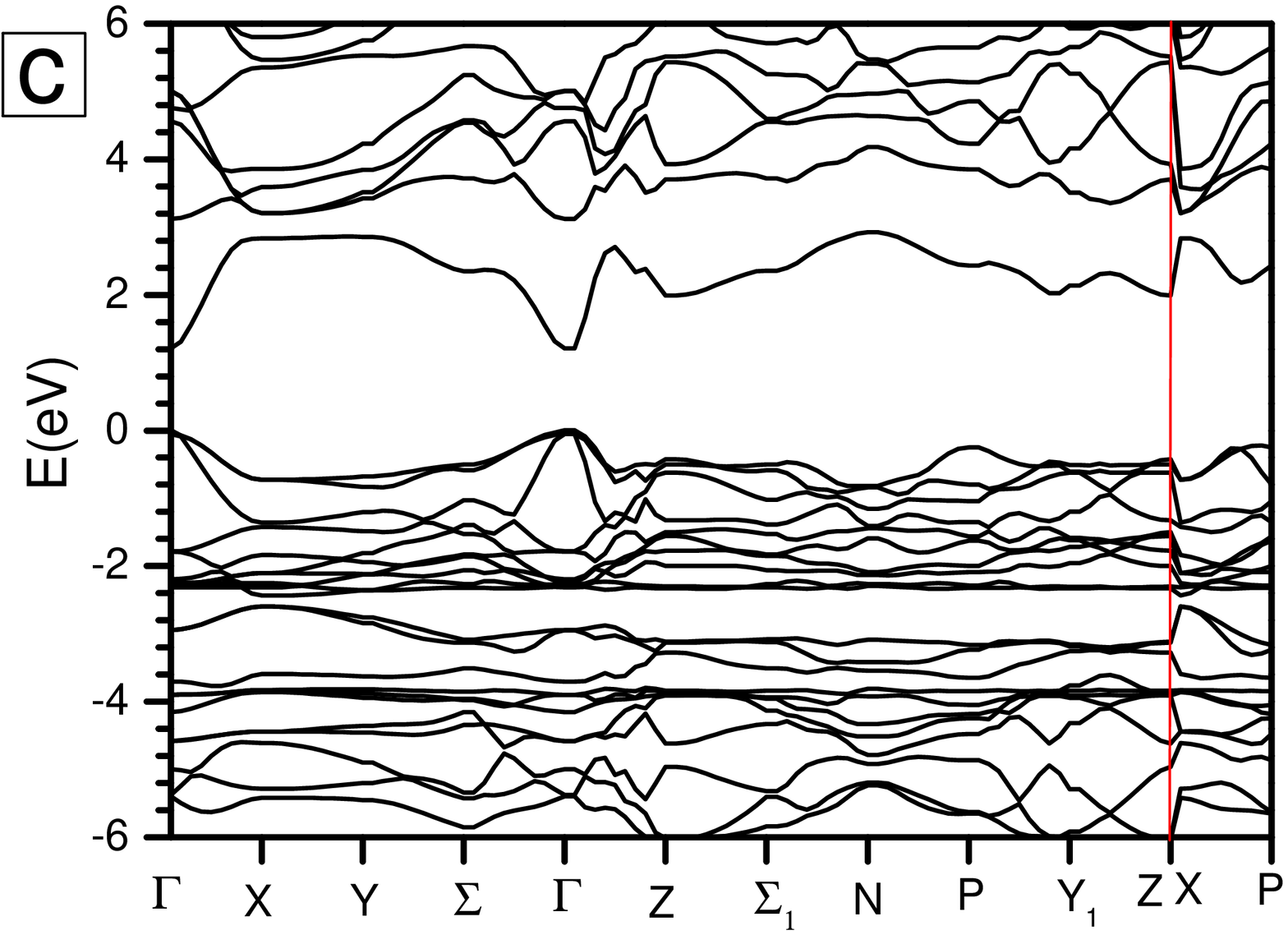}
\includegraphics[width=0.49\hsize]{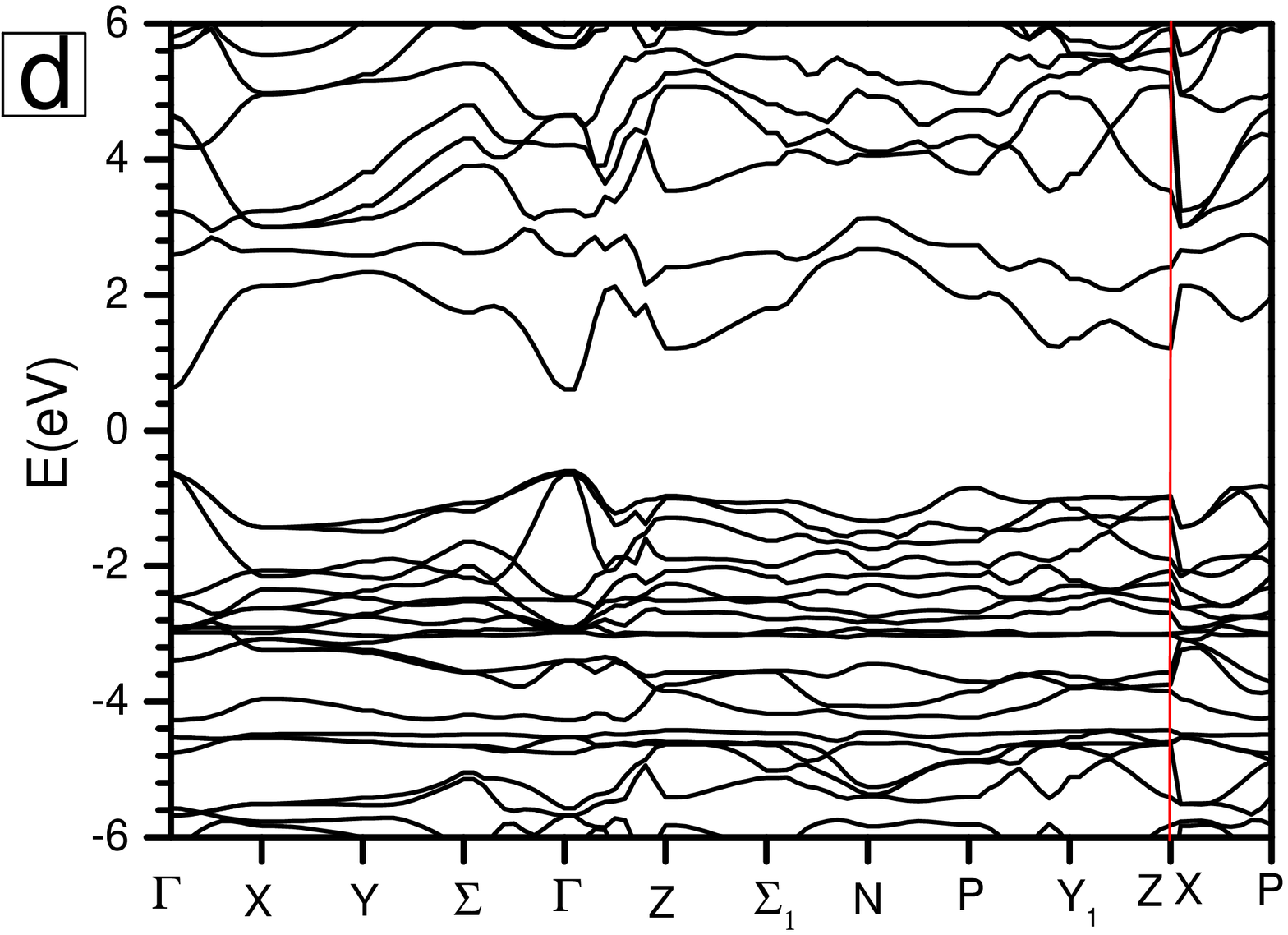}
\caption{\label{fig2} Band structure for (a) Cu$_2$ZnSnS$_4$, (b) Cu$_2$CdGeSe$_4$, (c)
Cu$_2$CdSnS$_4$, and (d) Cu$_2$HgGeS$_4$.}
\vspace{-0.5cm}
\end{center}
\end{figure}

\noindent
Two aspects of the low frequency behavior of the dielectric function are relevant to experiment.
One is the absorption edge, determined by onset of the imaginary part of the dielectric function.
The other one is zero frequency limit of the real part of the dielectric function, i.e.
the optical dielectric constant $\varepsilon_\infty$ which is given in table~\ref{table-S1}.\
\noindent
Table~\ref{table-S1} shows that in each family of chalcogenides, e.g. Cu$_2$HgGe-VI$_4$,
by replacing the element VI by one from the same group and with higher atomic number,
there is an increase in $\varepsilon_\infty$. Such an increase in $\varepsilon_\infty$ indicates
that the polarizability of the system tends to increase because of increasing ionicity of the bonds.
Replacing S with Se and then by Te (i.e. increasing the ionic radius) results in a more extended
electron cloud that screens the electric field more effectively and yields higher polarizability.
One can see that the plasma frequency ($\omega_p$) decreases with the same substitution. Given that the number of
valence electrons is the same for all of the studied chalcogenides, the decrease of $\omega_p$
can be understood as a consequence of the increased lattice constant on replacing an atom by a larger one.\

The energy of the first direct allowed transition (optical band gap) can be found from
the absorption spectra. Figs.~\ref{fig-S4}(a)-(c) present the arithmetic average of the absorption spectra. By replacing element VI by an element with a higher atomic number
there is a red shift in the absorption edge and the band gap also shows the same trend.
Moreover, from comparing the electronic band structure with the absorption spectrum we see that the first transition is direct and allowed.
The electronic band structure of four typical chalcogenides is shown in Figs.~\ref{fig2}(a)-(d).

\noindent
It is important for an absorber layer to highly absorb most part of the solar spectrum
and specifically the visible light. Substituting S by Se and then by Te (the VI element) increases
the maximum absorption of the Cu$_2$-II-IV-VI$_4$ compounds in the visible range (1.65 to 3.23 eV).
For example, the maximum absorption of Cu$_2$HgGeS$_4$, Cu$_2$HgGeSe$_4$, and Cu$_2$HgGeTe$_4$ is 2.21, 2.40,
and 4.17 $\times10^{-5}$ cm$^{-1}$, respectively. The opposite trend is found for the optical band gap.
It means that the optical band gap of Cu$_2$HgGeS$_4$, 1.21 eV is larger than that of Cu$_2$HgGeSe$_4$,
0.54 eV, and Cu$_2$HgGeTe$_4$, 0.38 eV.

\subsection{Power conversion efficiency}\label{optoelectronic}

\begin{figure}[!htp]
\begin{center}
\includegraphics[width=0.495\hsize]{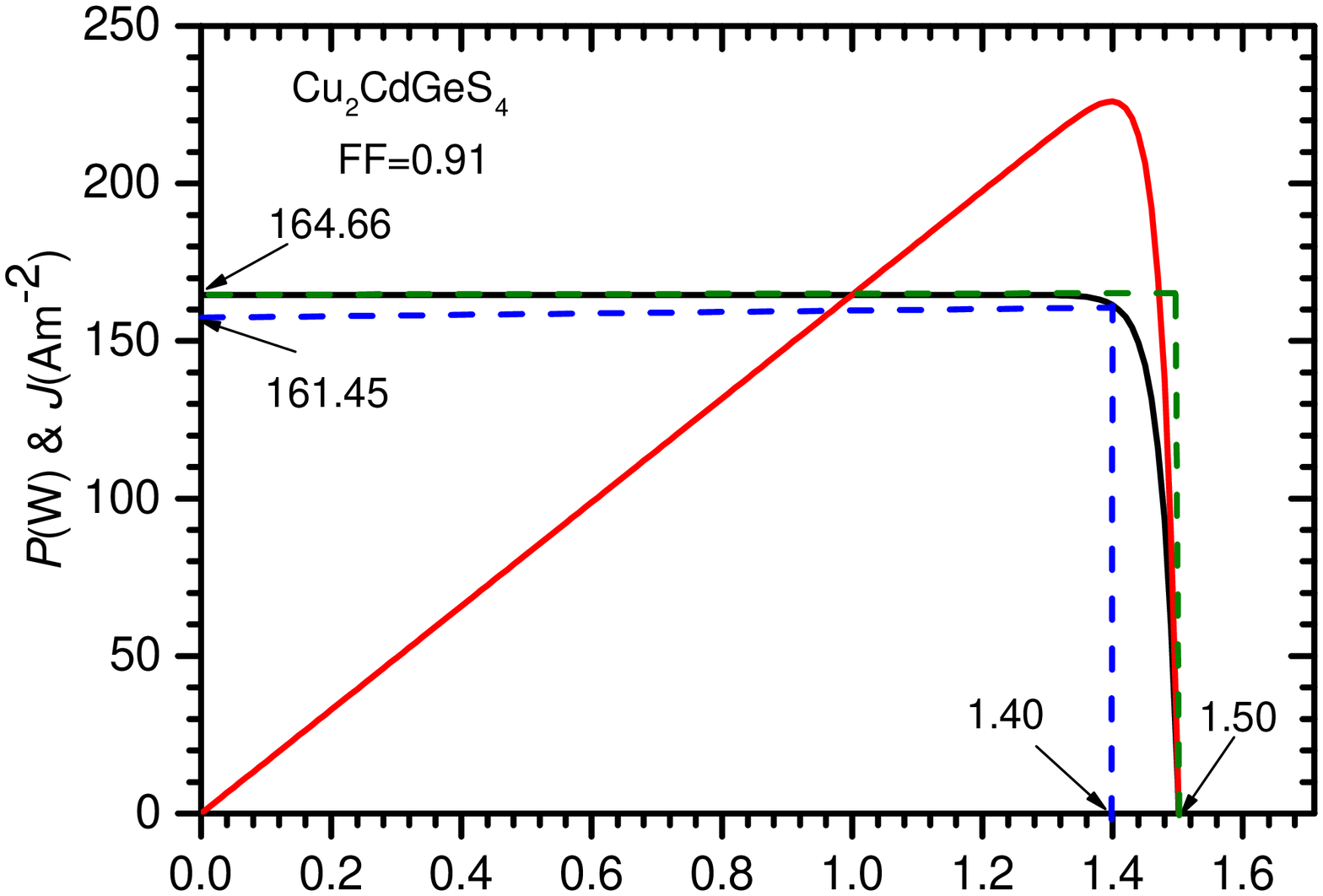}
\includegraphics[width=0.458\hsize]{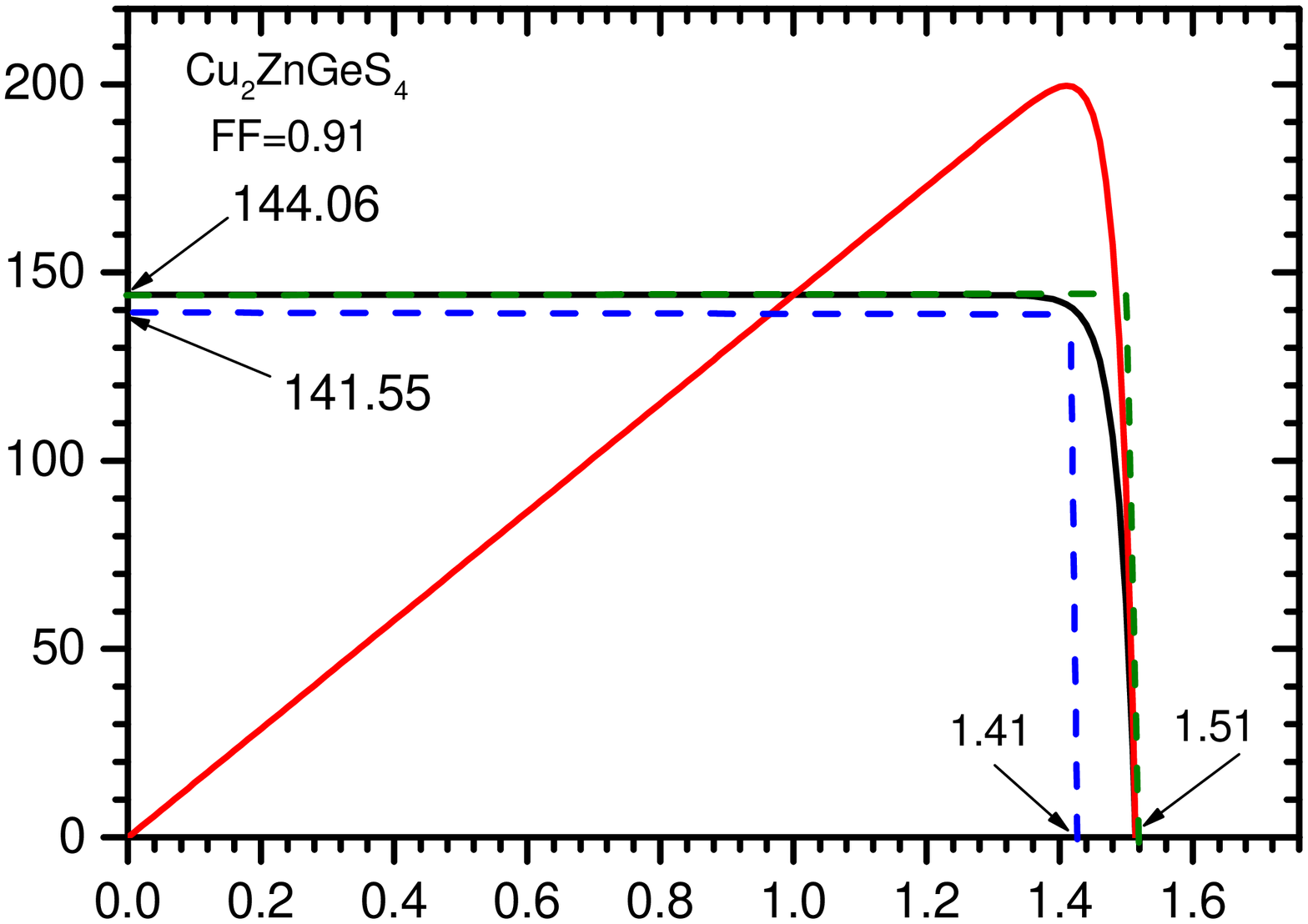}
\includegraphics[width=0.495\hsize]{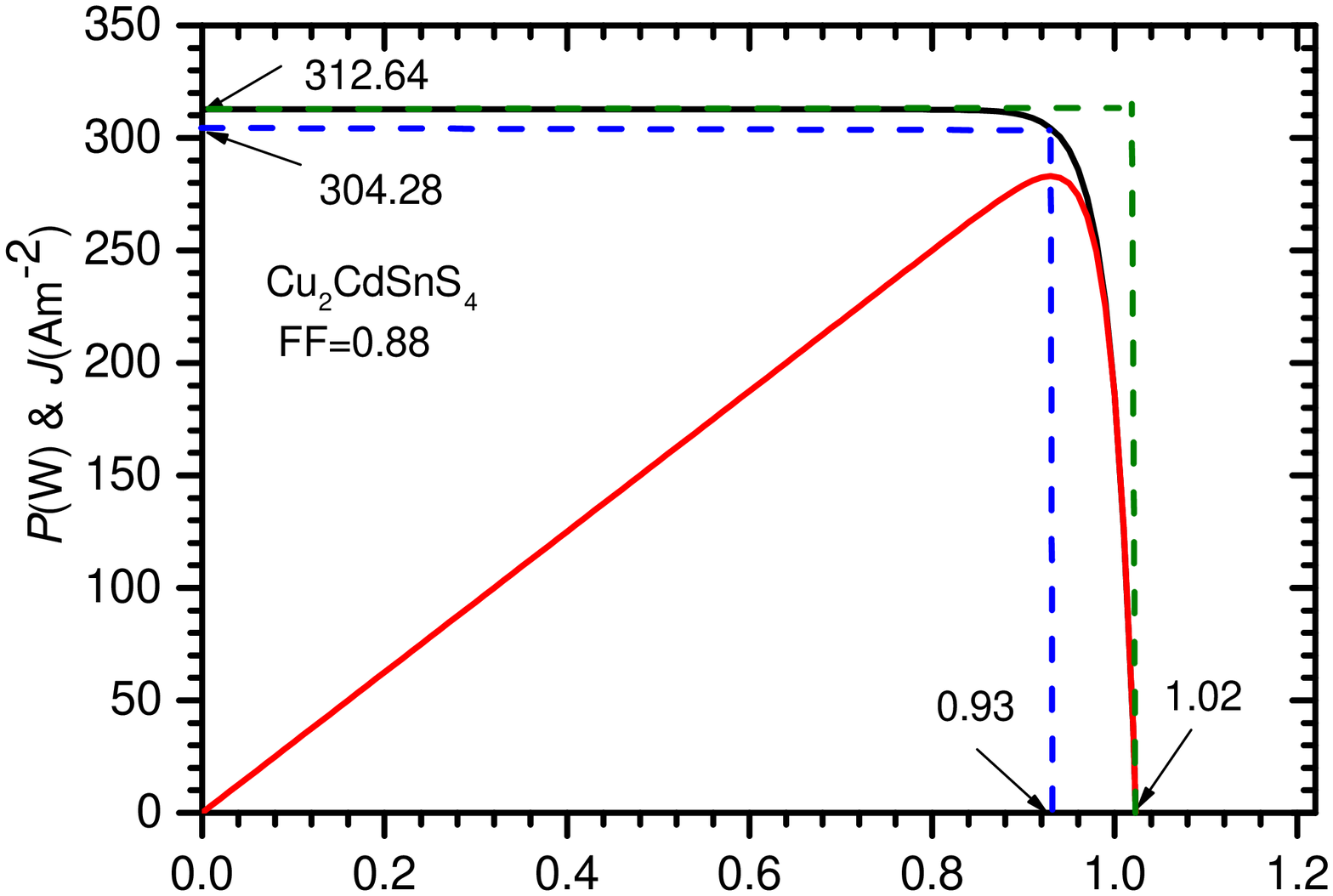}
\includegraphics[width=0.46\hsize]{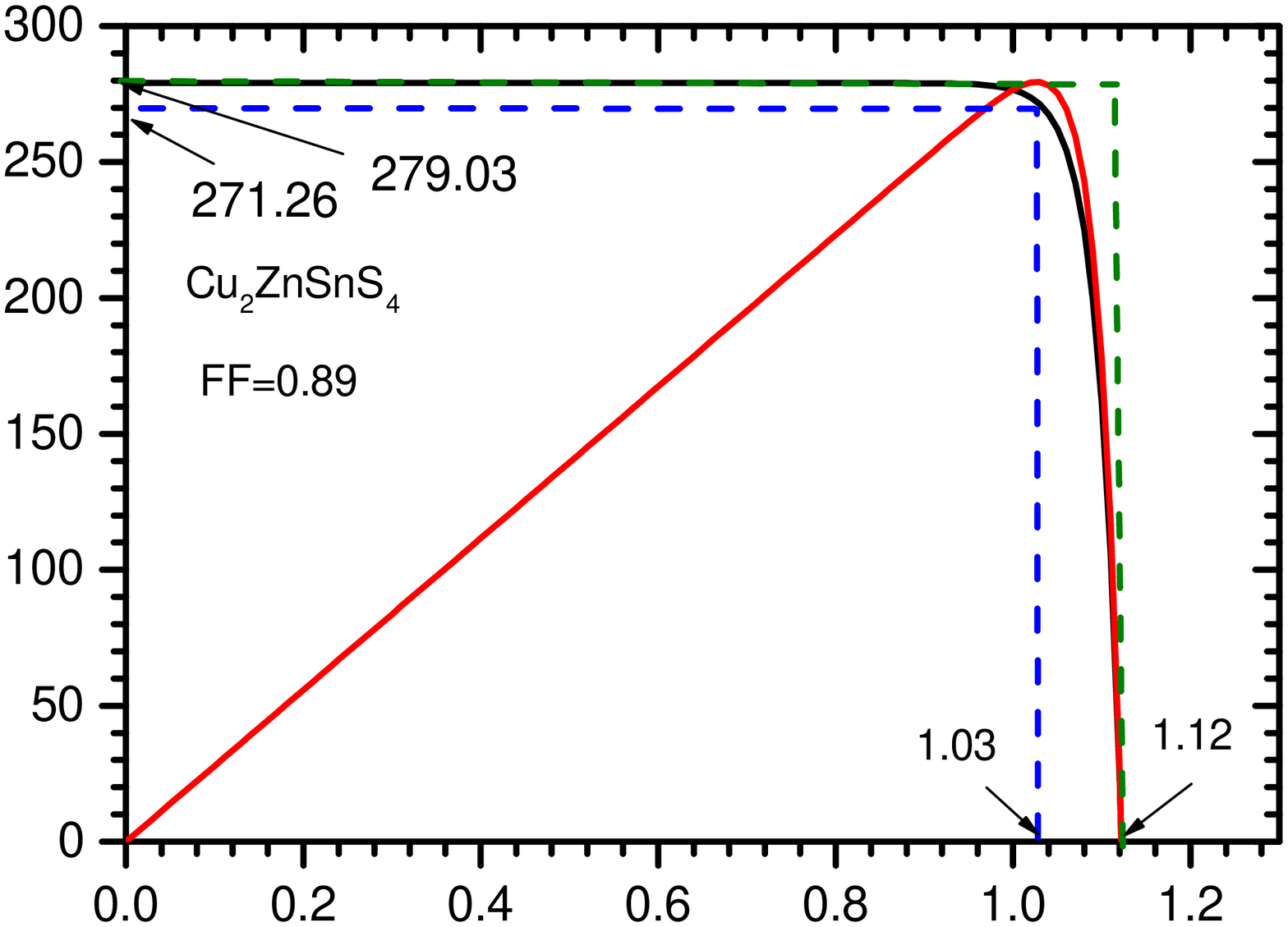}
\includegraphics[width=0.498\hsize]{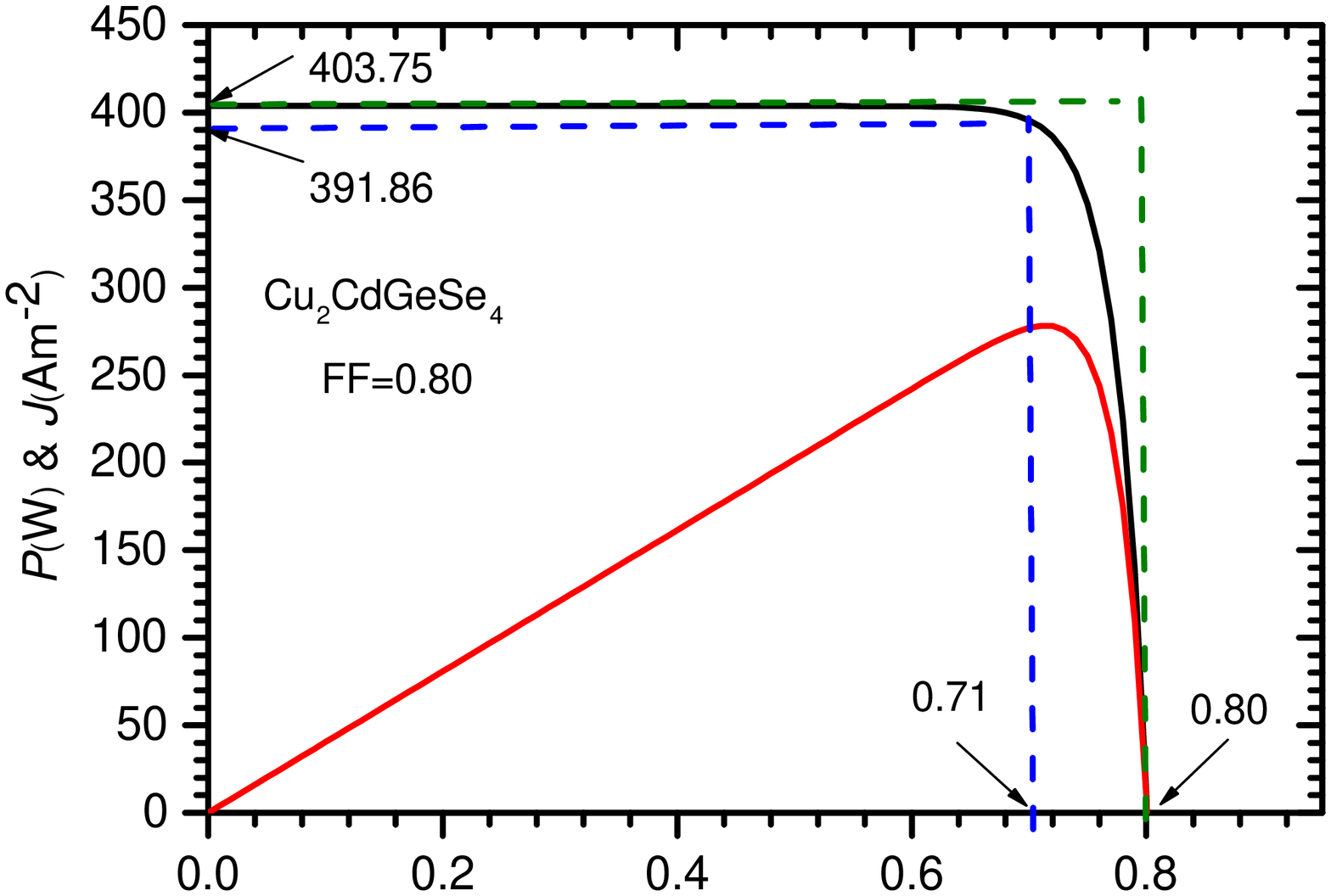}
\includegraphics[width=0.458\hsize]{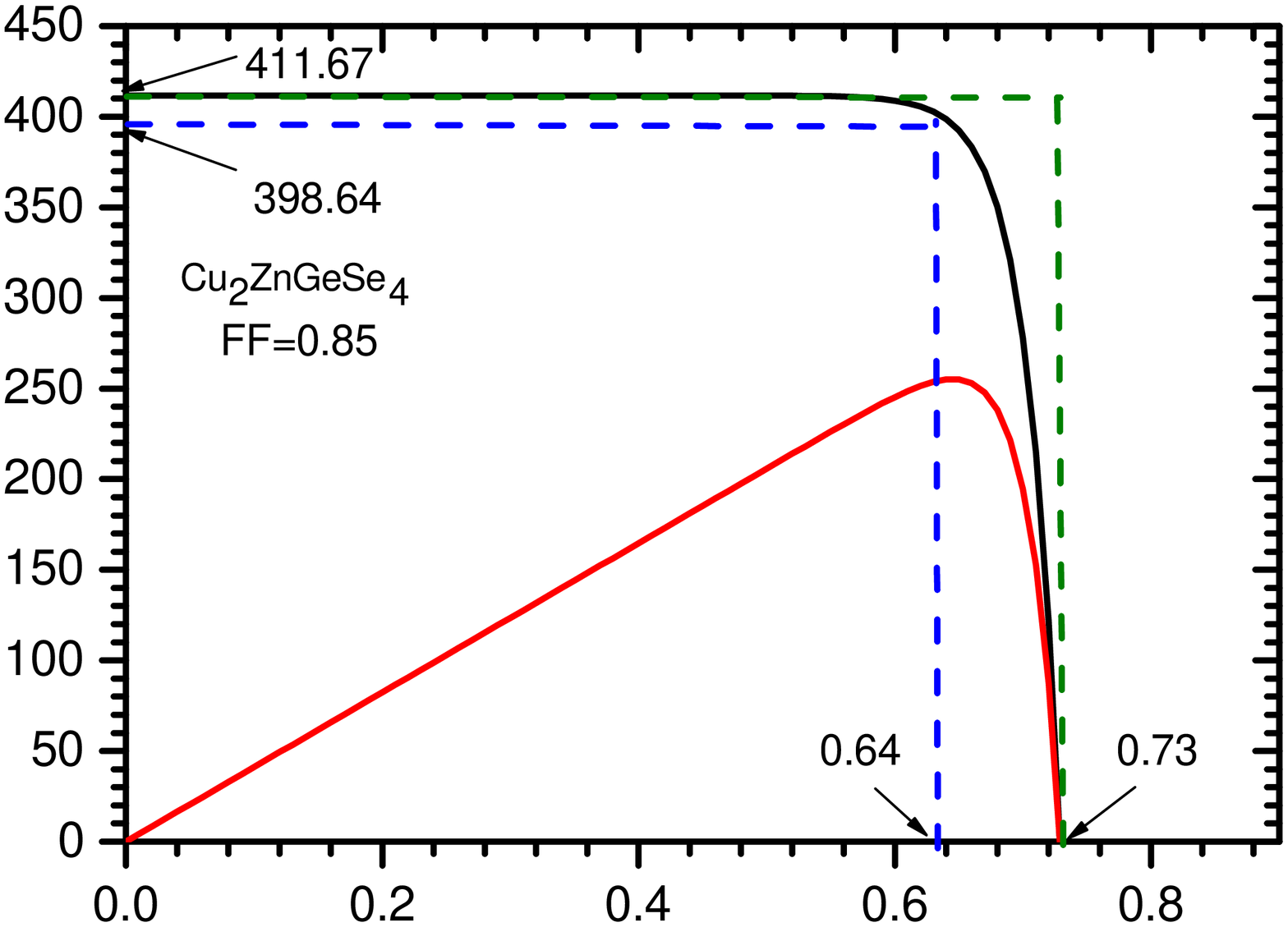}
\includegraphics[width=0.492\hsize]{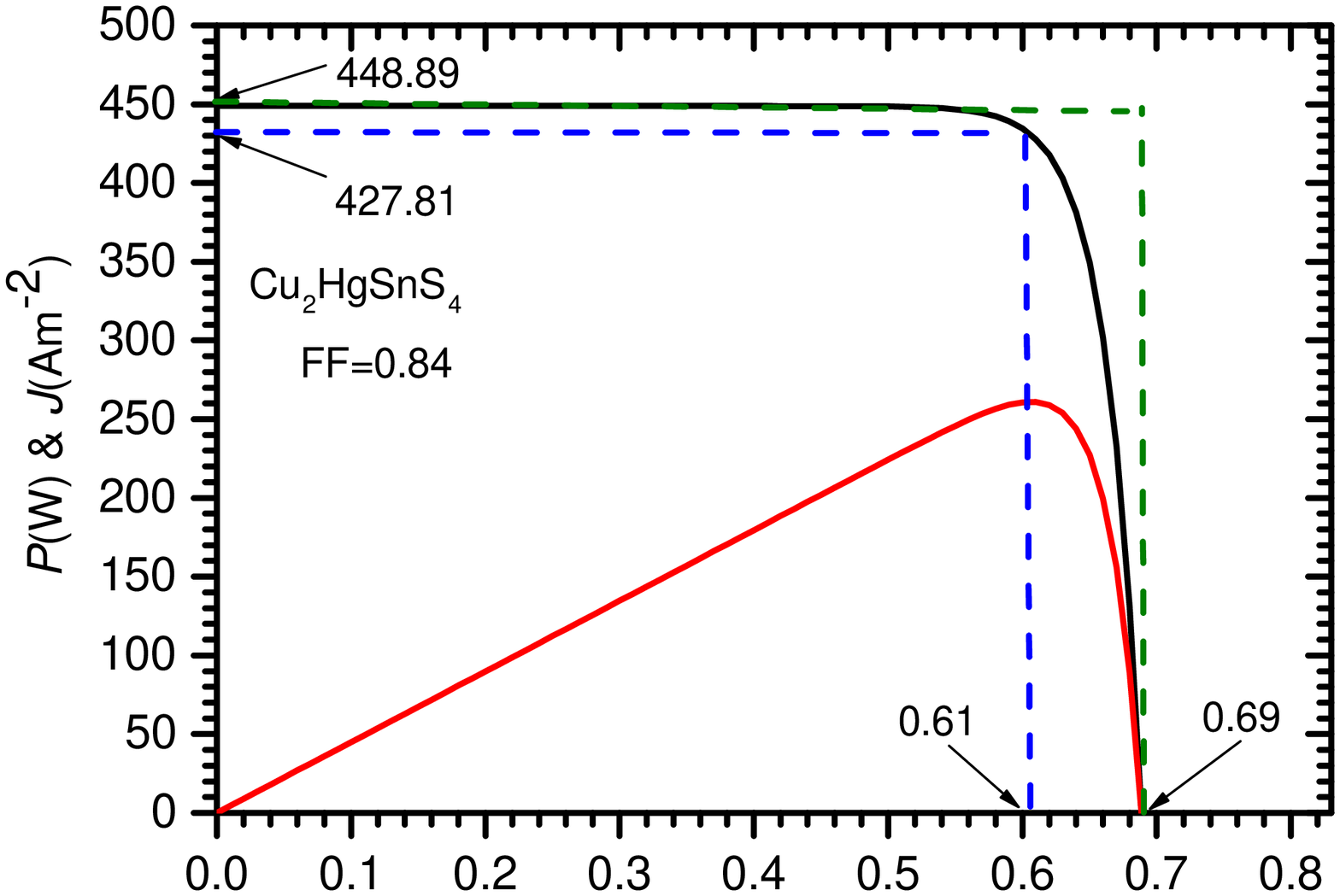}
\includegraphics[width=0.460\hsize]{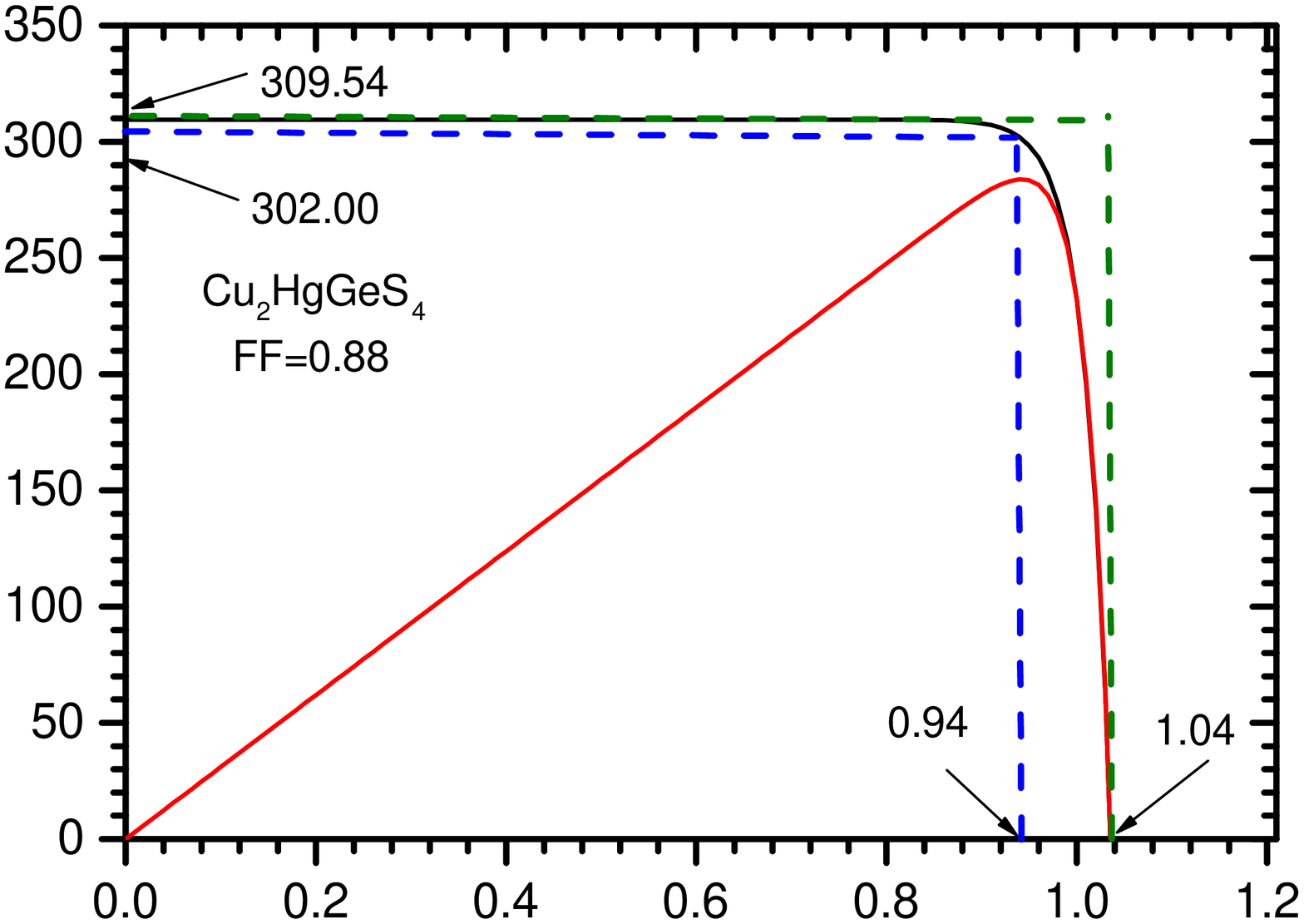}
\includegraphics[width=0.489\hsize]{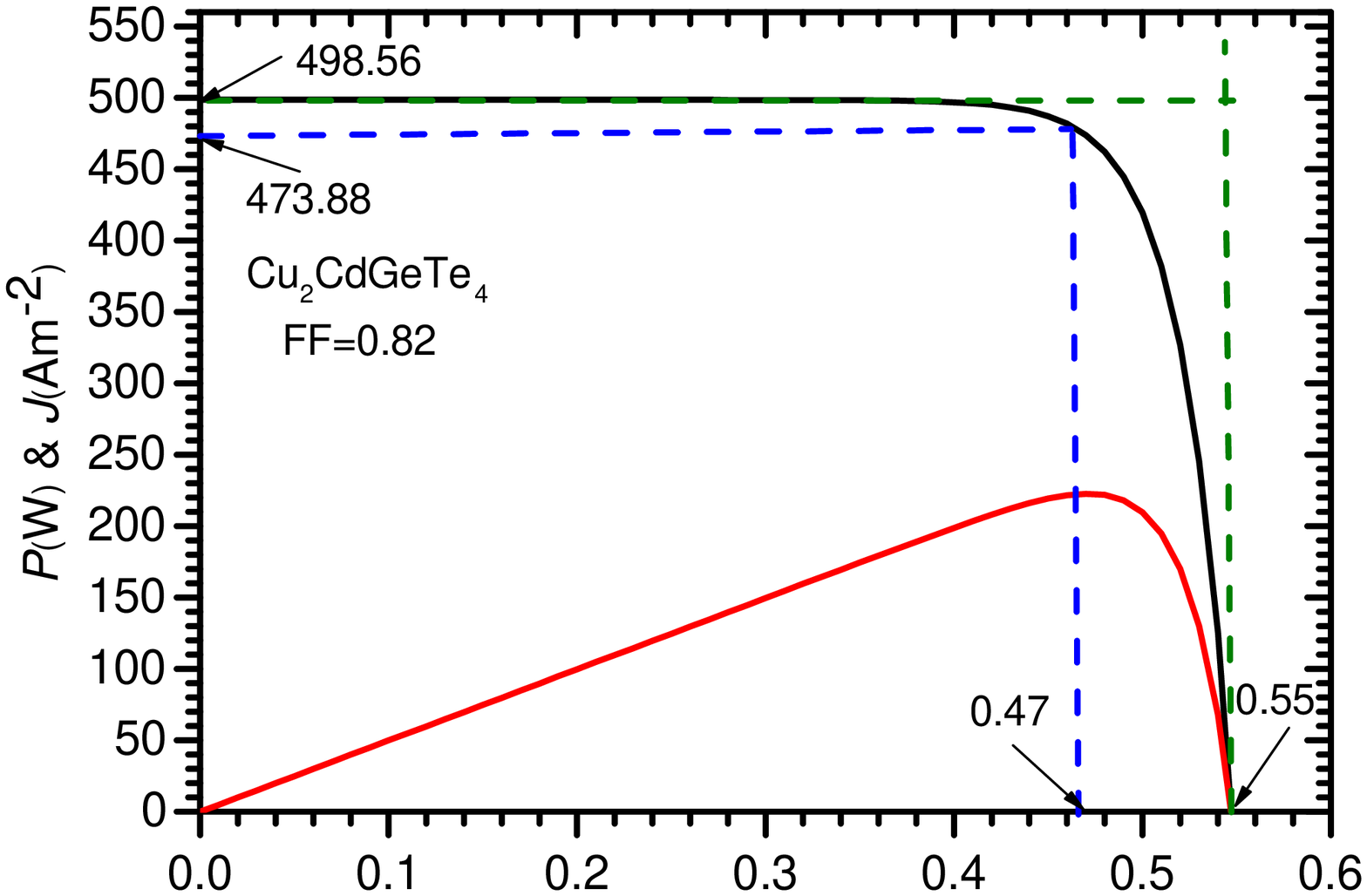}
\includegraphics[width=0.465\hsize]{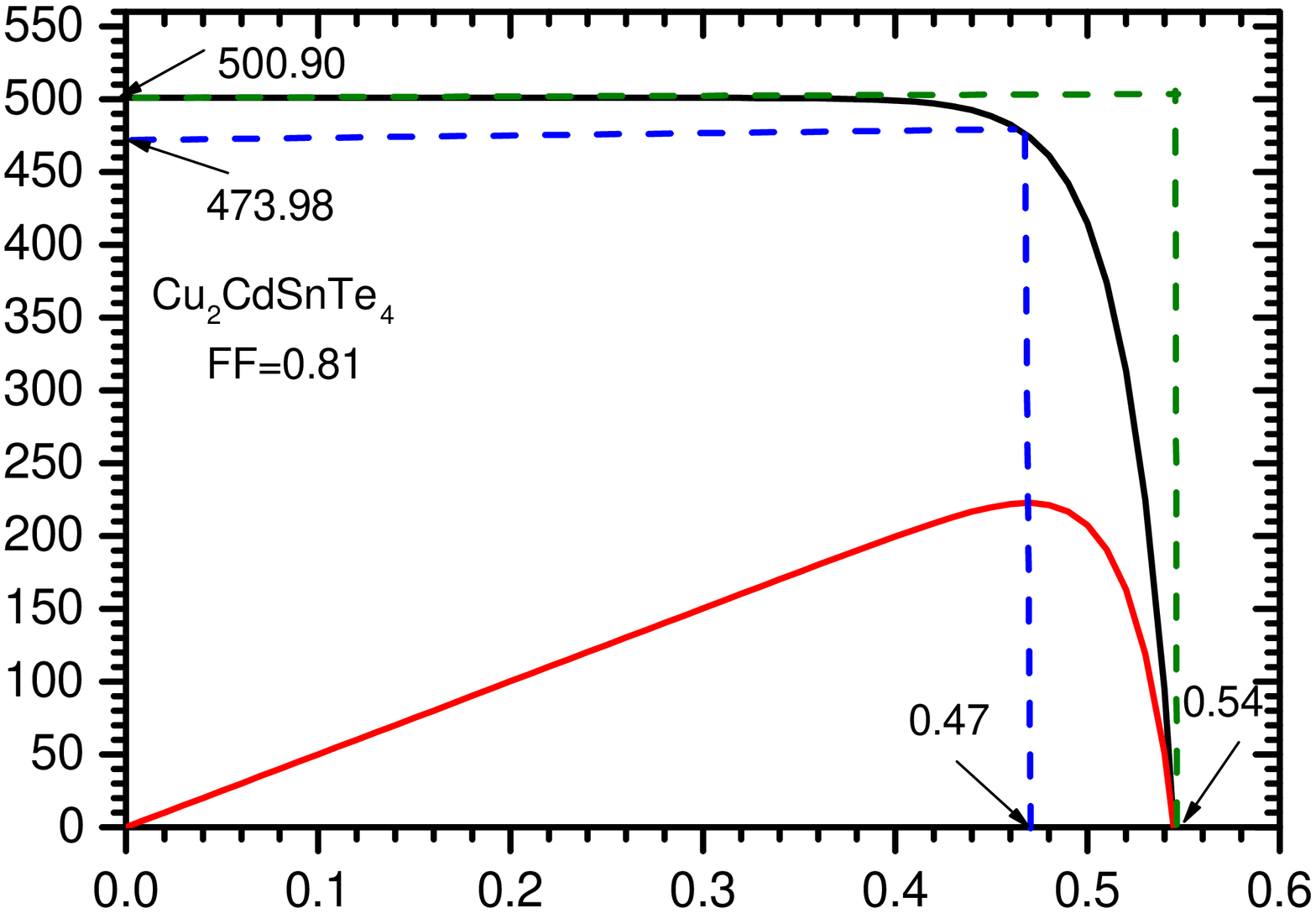}
\includegraphics[width=0.498\hsize]{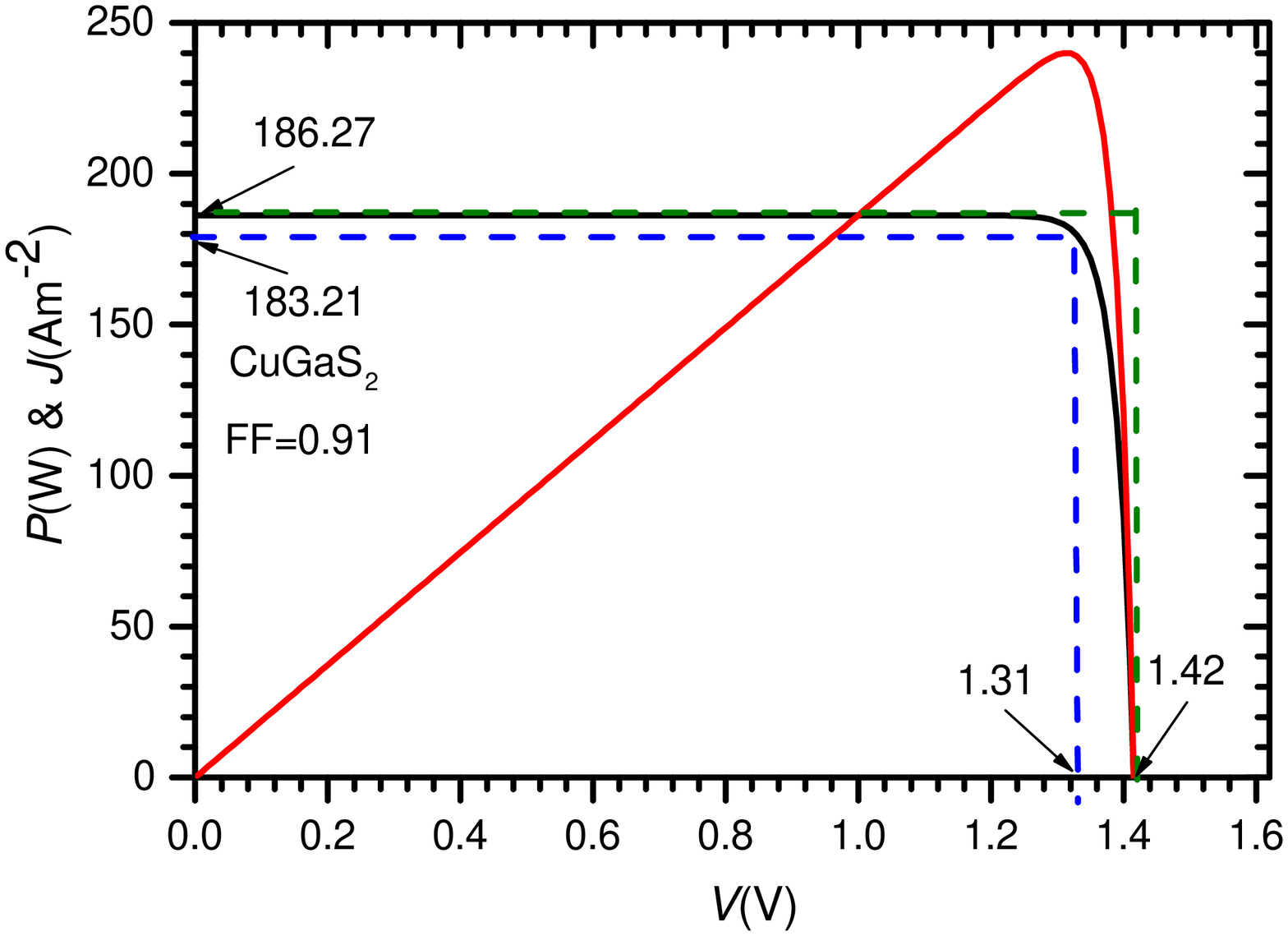}
\includegraphics[width=0.458\hsize]{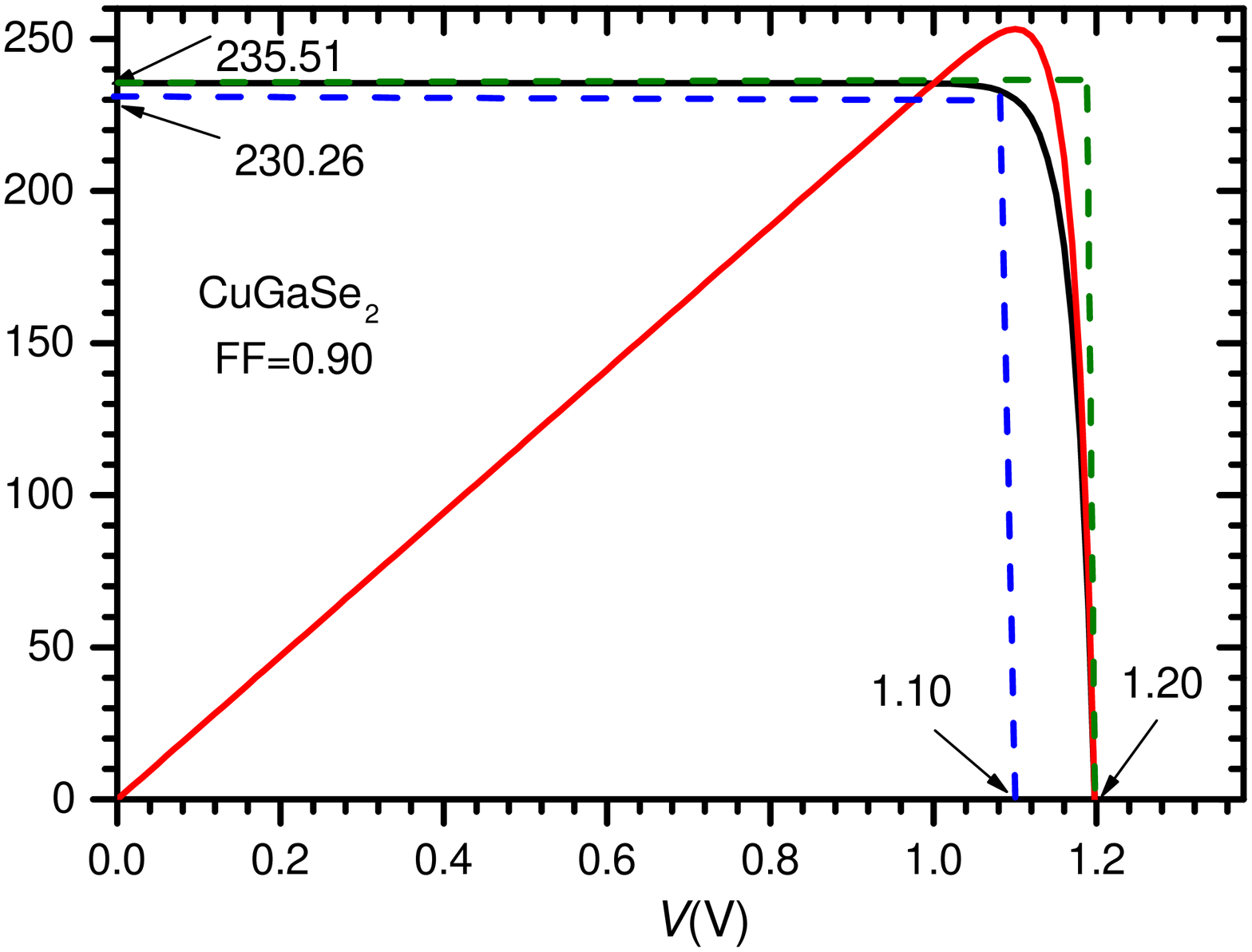}
\caption{current density (black curve) and power (red curve) of the absorber layer
(with the thickness of 0.5 $\mu$m) with respect to the voltage.
Blue, and green dashed lines represent the $P_m$ and, $P_{nominal}$, respectively. Lower and higher value
for voltages indicate $V_m$, and $V_{oc}$, respectively. Lower and higher value for current density values
show $J_m$, and $J_{sc}$, respectively.}
\label{fig3}
\vspace{-0.5cm}
\end{center}
\end{figure}

In order to calculate the SLME from the maximum output power of the absorber layer, we first plot
the current-voltage (J-V) and power-voltage (P-V) curve for the chalcogenides with an SLME below the corresponding SQ limit (See Fig.~\ref{fig3}).
As indicated later the SLME criterion can lead to anomalous results where the SLME value is beyond the SQ limit. We therefore limit ourselves to those compounds with an SLME value below the SQ limit.\\
\noindent
The same plot is shown for two common thin-film solar cell materials, CuGaS$_2$, and CuGaSe$_2$.
The voltage is the difference between the quasi Fermi level for electrons and holes.
This value can be changed by applying an electric field (i.e. via incident photons on
the solar cell). The voltage changes between zero and its maximum value which is $E_g/e$ of the material.
$P_m$ and $P_{nominal}$ are presented by blue and green dashed lines, respectively.
Figure~\ref{fig3} also gives $V_m$, and $V_{oc}$.
For each compound, the lower voltage is $V_m$, and the higher one corresponds to
$V_{oc}$. Each plot gives two more values, namely $J_m$ (lower one) and $J_{sc}$ (higher one).
According to the definition of $J_{sc}$ and $J_m$, $J_{sc}$ is always larger than $J_m$ and
this difference depends on the recombination rate.
Likewise, Eq.~\ref{eq4} implies that $V_{oc}$ is always larger than $V_m$.
Altogether, $P_{nominal}$ is always larger by a factor (FF) than $P_m$. The FF values are given
for each compound in the corresponding plot in Fig~\ref{fig3}.
The $J$-$V$ plot shows a large $J_{sc}$ of 411.67 Am$^{-2}$ for Cu$_2$ZnGeSe$_4$.
However, this compound has a small $V_{oc}$ compared with the others that results in low output power,
FF, and efficiency. On the other hand, Cu$_2$CdGeS$_4$ with low $J_{sc}$ has a
large $V_{oc}$ that yields a large efficiency and largest FF.

%figure3*

\begin{figure}[!htp]
\begin{center}
\includegraphics[width=0.8\hsize]{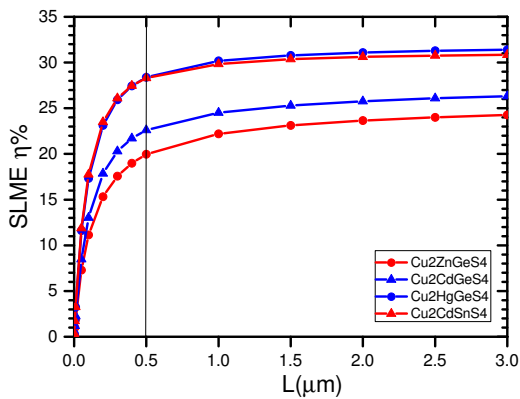}
\caption{\label{fig4} The SLME as a function of thin film thickness for Cu$_2$CdGeS$_4$,
Cu$_2$ZnGeS$_4$, Cu$_2$HgGeS$_4$, and Cu$_2$CdSnS$_4$.The vertical line indicates the thickness used in
Fig.~\ref{fig5}}
\vspace{-0.5cm}
\end{center}
\end{figure}

Typical absorber layers (e.g. Cu\{In,Ga\}Se$_2$) are a few micrometers thick, further decrease in thickness
is desirable to reduce processing times and material usage \cite{micron_CIGS}. However, there is a limit on
the thickness of the absorber layer based on the desirable efficiency value. The SLME depends on the film
thickness and it increases as $L$ increases. It converges to the corresponding SQ limit for very large thicknesses of the considered film. In fact, the materials with the same band gap might have different rate of the convergence because of their different optical properties. Finding a material that converges fast to the SQ limit is desirable. Figure~\ref{fig4} represents the SLME as a function of thin film thickness for four studied chalcogenides. On the one hand, there is a slight increase in the efficiency by increasing the film thickness for values larger than 0.5 $\mu$m.  On the other hand,
by making the film thinner than 0.5 $\mu$m the SLME diminishes. This means that we cannot make
the absorber layer thinner than 0.5 $\mu$m without loosing too much of its efficiency. Therefore,
we use $L = 0.5$ $\mu$m for the thickness of the studied absorber layers.

Figure~\ref{fig4} shows two materials, Cu$_2$CdSnS$_4$ and Cu$_2$HgGeS$_4$ with almost the same band gap
resulting in very similar efficiencies within the SQ limit. In the range of sub-micrometer thicknesses,
the difference in the SLME values mainly results from the different corresponding optical properties.\

\begin{figure}[!htp]
\begin{center}
\includegraphics[width=0.8\hsize]{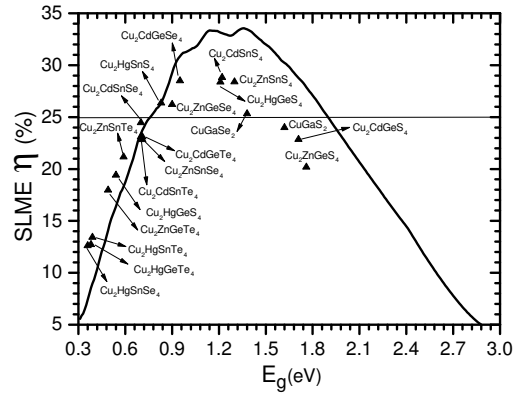}
\caption{SLME versus the band gap (E$_g$) for the studied chalcogenides at $L$ = 0.5 $\mu$m.
The full line presents the SQ limit.}
\vspace{-0.8cm}
\label{fig5}
\end{center}
\end{figure}

Figure~\ref{fig5} presents the SLME parameter for the studied compounds as a function of
the band gap energy. The calculated SLME value for two well-known absorber layer materials (CuGaS$_2$, and CuGaSe$_2$) is also represented. The SQ limit as the upper limit on the efficiency is also given as a curve.\

%figure4*

%figure5*

\noindent
One can see from Fig.~\ref{fig5} that the SLME value of some of the considered compounds is higher than the corresponding SQ limit, which is a theoretical upper limit for the efficiency of the absorber layer. In a separate publication we analyze this anomaly in more detail \cite{Marnik-CuAu}. Nevertheless, as shown by Yu and Zunger \cite{zunger} the SLME is more powerful than the SQ limit in ranking the compounds based on their power conversion efficiency by including the optical properties of the materials. For example a comparison between Cu$_2$CdGeSe$_4$ and Cu$_2$ZnSnS$_4$ shows that it is possible to have a high efficiency absorber layer with a non-optimum band gap material. Cu$_2$CdGeSe$_4$ has a band gap of 0.95 eV and the band gap of Cu$_2$ZnSnS$_4$ is 1.30 eV. According to the SQ limit, the latter is expected to have a higher efficiency. However, the former compound has a higher absorption that results in a higher efficiency.\\
\noindent
Nevertheless, SLME is more powerful than SQ limit in ranking the compounds based on their power conversion efficiency by including the optical properties of the materials. This fact can be understood by making a comparison between the studied compounds. For example a comparison between Cu$_2$CdGeSe$_4$ and Cu$_2$ZnSnS$_4$ shows that it is possible to have a high efficiency absorber layer with a non-optimum band gap material. Cu$_2$CdGeSe$_4$ has a band gap of 0.95 eV and the band gap of
Cu$_2$ZnSnS$_4$ is 1.30 eV. According to the SQ limit, the latter is expected to
have a higher efficiency. However, the former compound has a higher absorption that results in
a higher efficiency.\\
\indent
Based on Fig.~\ref{fig5} our calculations identify four quaternary Cu-based chalcogenides, namely Cu$_2$CdGeSe$_4$, Cu$_2$CdSnS$_4$, Cu$_2$HgGeS$_4$, and Cu$_2$ZnSnS$_4$, as possible absorber layers with a power efficiency higher than 25\%.
Their theoretical efficiency stays considerably above the other studied chalcogenides. The latter identified material is already used as an absorber layer in photovoltaic cells.
Its SLME value differs by 8\% from that of Ref.~\cite{perovskite}, where the absorption spectrum was calculated using the GGA functional. It is already proven that
HSE06 provides a dielectric function in much better agreement with experiment than GGA or
LDA functionals \cite{dielectric_exp_PBE_HSE1,dielectric_exp_PBE_HSE2}.
The measured cell efficiency for CZTS is 12.6\% lower than its SLME value \cite{CZTS_12.6} because of (i) the difference between solar cell and absorber layer and (ii) the presence of electron--hole recombination centers. The latter should be prohibited to get the maximum theoretical power conversion efficiency.
Further studies on the formation of native defects in the identified compounds are required to
understand better the power conversion efficiency limit applied to these materials.\\
\indent
We also calculate the band structure and optical properties for CuGaS$_2$, and CuGaSe$_2$.
The HSE06 calculations result in 1.62, and 1.38 eV band gap for CuGaS$_2$, and CuGaSe$_2$, respectively.
There is an 33\%, and 17\% underestimation with respect to the experiment band gap for CuGaS$_2$, and CuGaSe$_2$, respectively \cite{12L}.
We calculate 24\%, and 25.33\% for the SLME of the considered compounds. A comparison between our results with existing results that used the GW approximation for the band gap calculation \cite{zunger} and HSE for the optical properties shows 7.5\% and 1.3\% difference in the SLME value for these compounds.
However, both calculations result in a higher band gap and a higher SLME for CuGaSe$_2$.

\section{\label{HT_summary}Conclusion\protect\\}

The results of the present work show that the optoelectronic properties of the studied
Cu-based chalcogenides Cu$_2$-II-IV-VI$_4$ strongly depend on the element VI in
the composition of the material. The change of element VI has a higher effect than
changing element II, or IV in altering the characteristics of the studied
chalcogenides. Replacement of the element VI by one from the same group with a higher atomic number
decreases the plasma frequency and band gap and at the same time results in an increase of the lattice
parameters, and optical dielectric constant. A clear red shift in the absorption edge is observed which is
correlated with the decrease in the band gap.
Further studies of the compounds of interest show that besides the fundamental band gap that plays a
main role in the efficiency of an absorber layer, the absorption coefficient is an essential characteristic.
The absorption coefficient is important to compare the efficiency of two compounds with the same band gap.
In that case two compounds have the same SQ efficiency, but the one with
higher absorptivity has a higher SLME value.
Finally, the results of the calculations identify Cu$_2$II-GeSe$_4$ with II=Cd, and Hg and Cu$_2$-II-SnS$_4$
with II=Cd, and Zn as high efficiency absorber layers.

\section{\label{Acknowledgements}Acknowledgements\protect\\}

We acknowledge the financial support from the FWO-Vlaanderen through project G.0150.13N and a GOA fund from the University of Antwerp. The computational resources and services used in this work were provided by
the VSC (Flemish Supercomputer Center) and the HPC infrastructure of the University of Antwerp (CalcUA),
both funded by the FWO-Vlaanderen and the Flemish Government--department EWI.

\appendix
\setcounter{figure}{0} \renewcommand{\thefigure}{A.\arabic{figure}}

\section{}\label{appendix}

Table~\ref{table-S1} presents the list of studied stannite Cu-based chalcogenides in the following order:
Cu$_2$Zn-based, Cu$_2$Cd-based, and then Cu$_2$Hg-based compounds.
For each compound, the first row presents the calculated HSE06 results,
and the following rows contain the available experimental and theoretical data.

\renewcommand{\arraystretch}{1}
\begin{longtable*}{p{0.12\textwidth}p{0.10\textwidth}p{0.09\textwidth}p{0.09\textwidth}p{0.09\textwidth}p{0.10\textwidth}p{0.18\textwidth}}
\caption{\ HSE06 calculated lattice constants ($a$ and $c$ in \AA), band gap ($E_g$ in eV),
enthalpy of formation ($\triangle H_f$ in eV)$^{i}$, plasma frequency ($\omega_p^{avr.}$ in eV)$^{ii}$, and optical dielectric constant ($\varepsilon_\infty^{avr.}$)$^{ii}$ for the studied Cu-based chalcogenides.
The calculated data are compared with other available results in the literature.\label{table-S1}}\\
\hline
\multicolumn{1}{l}{Compound} & \multicolumn{1}{l}{a} & \multicolumn{1}{l}{c/a} & \multicolumn{1}{l}{E$_g$} & \multicolumn{1}{l}{$\Delta$H$_f$} & \multicolumn{1}{l}{$\omega_p^{avr.}$} & \multicolumn{1}{l}{$\varepsilon_\infty^{avr.}$ ($\varepsilon_\infty^{\bot}$, $\varepsilon_\infty^{\|}$)}\\
\hline
\endfirsthead
\multicolumn{7}{l}{Table I continued}\\
\hline
\multicolumn{1}{l}{Compound} & \multicolumn{1}{l}{a} & \multicolumn{1}{l}{c/a} & \multicolumn{1}{l}{E$_g$} & \multicolumn{1}{l}{$\Delta$H$_f$} & \multicolumn{1}{l}{$\omega_p^{avr.}$} & \multicolumn{1}{l}{$\varepsilon_\infty^{avr.}$ ($\varepsilon_\infty^{\bot}$, $\varepsilon_\infty^{\|}$)}\\
\hline
\endhead
\endfoot
\endlastfoot

Cu$_2$ZnGeS$_4$ & 5.30 & 2.02 & 1.76 & -2.99 & 19.62 & 6.09 (5.96, 6.37)\\
& 5.34$^{iii}$\cite{Cu2ZnGeTe4_exp} & 1.97$^{iii}$\cite{Cu2ZnGeTe4_exp} & 2.04$^{iii}$\cite{Cu2ZnGeTe4_the} & & &\\
& 5.33$^{iv}$\cite{Cu2ZnGeTe4_the} & 2.06$^{iv}$\cite{Cu2ZnGeTe4_the} & 2.14$^{vi}$\cite{Cu2ZnGeTe4_the}
& & & 6.8$^{iv}$\cite{Cu2ZnGeTe4_the}\\
&&&&&&\\

Cu$_2$ZnGeSe$_4$ & 5.60 & 2.01 & 0.90 & -2.31 & 18.16 & 7.56 (7.36, 7.97)\\
& 5.63$^{iii}$\cite{117} & 1.96$^{iii}$\cite{117} & 1.29$^{iii}$\cite{Cu2ZnGeTe4_the} & & & \\
& 5.38$^{iv}$\cite{Cu2ZnGeTe4_the} & 2.02$^{iv}$\cite{Cu2ZnGeTe4_the} & 1.32$^{vi}$\cite{Cu2ZnGeTe4_the}
& & & 9.01$^{iv}$\cite{Cu2ZnGeTe4_the}\\
&&&&&&\\

Cu$_2$ZnGeTe$_4$ & 6.04 & 1.99 & 0.49 & -3.21 & 16.53 & 10.17 (9.89, 10.73)\\
& 5.60$^{iii}$\cite{Cu2ZnGeTe4_exp} & 1.99$^{iii}$\cite{Cu2ZnGeTe4_exp}& & & & \\
& 6.09$^{iv}$\cite{Cu2ZnGeTe4_the} & 2.00$^{iv}$\cite{Cu2ZnGeTe4_the} & 0.55$^{vi}$\cite{Cu2ZnGeTe4_the}
& & 17.93$^{iv}$\cite{Cu2ZnGeTe4_the}&\\
&&&&&&\\

Cu$_2$ZnSnS$_4$ & 5.42 & 2.01 & 1.30 & -3.15 & 19.00 & 6.25 (6.12, 6.53)\\
& 5.44$^{iii}$\cite{pv8}& 2.01$^{iii}$\cite{pv8} & 1.29$^{iii}$\cite{pv8} & & & 6.48$^{iii}$\cite{pv8}\\
& 5.34$^{v}$\cite{123} & 2.01$^{v}$\cite{123} & 1.27$^{v}$\cite{123} & & & 6.99$^{v}$(z)\cite{123}\\
&&&&&&\\

Cu$_2$ZnSnSe$_4$ & 5.71 & 2.00 & 0.71 & -4.23 & 16.37 & 7.74 (7.56, 8.12)\\
& 5.61$^{iii}$\cite{Cu2ZnSnTe4} & 1.99$^{iii}$\cite{Cu2ZnSnTe4} & 1.41$^{iii}$\cite{120} & & &\\
& 5.61$^{v}$\cite{123} & 1.99$^{v}$\cite{123} & 0.69$^{v}$\cite{123}& & & 8.19(x), 8.27(z)$^{v}$\cite{123}\\
&&&&&&\\

Cu$_2$ZnSnTe$_4$ & 6.13 & 2.00 & 0.58 & -2.24 & 16.38 & 9.74 (9.53, 10.17)\\
& 6.20$^{iii}$\cite{Cu2ZnSnTe4} & 1.99$^{iii}$\cite{Cu2ZnSnTe4} & 0.5$^{iii}$\cite{Cu2ZnSnTe4} & &
14$^{iii}$\cite{Cu2ZnSnTe4} & \\
& 6.20$^{iv}$\cite{Cu2ZnSnTe4} & 1.99$^{iv}$\cite{Cu2ZnSnTe4} & & & &\\
&&&&&&\\

Cu$_2$CdGeS$_4$ & 5.52 & 1.91 & 1.71 & -3.97 & 19.02 & 6.06 (6.24, 7.88)\\
& 5.34$^{iii}$\cite{Cu2CdGeS4} & 1.97$^{iii}$\cite{Cu2CdGeS4} & & & & \\

Cu$_2$CdGeSe$_4$ & 5.79 & 1.92 & 0.95 & -3.08 & 18.46 & 7.51 (7.33, 7.88)\\
& 5.75$^{iii}$\cite{136} & 1.92$^{iii}$\cite{136} & 1.20$^{iii}$\cite{136} & & &\\
&&&&&&\\

Cu$_2$CdGeTe$_4$ & 6.20 & 1.93 & 0.71 & -2.24 & 16.36 & 9.78 (9.50, 10.33)\\
& 6.13$^{iii}$\cite{130} & 1.94$^{iii}$\cite{130} &  &  & & \\
&&&&&&\\

Cu$_2$CdSnS$_4$ & 5.62 & 1.940 & 1.22 & -3.74 & 18.68 & 6.25 (6.10, 6.56)\\
& 5.59$^{iii}$\cite{Cu2CdSnS4} & 1.94$^{iii}$\cite{Cu2CdSnS4} & 1.45$^{iii}$\cite{140}& & & \\
&&&&&&\\

Cu$_2$CdSnSe$_4$ & 5.88 & 1.95 & 0.70 & -4.01 & 17.74 & 7.72 (7.47, 8.23)\\
& 5.81$^{iii}$\cite{stannite1} & 1.97$^{iii}$\cite{stannite1} & 0.96$^{iii}$\cite{144} & & & \\
&&&&&&\\

Cu$_2$CdSnTe$_4$ & 6.27 & 1.97 & 0.70 & -2.28 & 16.19 & 9.44 (9.16, 9.99)\\
& 6.20$^{iii}$\cite{Cu2CdSnTe4} & 1.98$^{iii}$\cite{Cu2CdSnTe4} & & & & \\
&&&&&&\\

Cu$_2$HgGeS$_4$ & 5.52 & 1.92 & 1.21 & -3.05 & 19.43 & 6.79 (6.75, 6.87)\\
& 5.49$^{iii}$\cite{stannite2} & 1.92$^{iii}$\cite{stannite2} &&&&\\
&&&&&&\\

Cu$_2$HgGeSe$_4$ & 5.79 & 1.92 & 0.54 & -3.14 & 18.40 & 8.90 (8.74, 9.23)\\
& 5.69$^{iii}$\cite{stannite1} & 1.93$^{iii}$\cite{stannite1} &&&&\\

Cu$_2$HgGeTe$_4$ & 6.19 & 1.94 & 0.38 & -1.74 & 17.06 & 11.93 (11.81, 12.18)\\
& 6.11$^{iii}$\cite{130} & 1.95$^{iii}$\cite{130} & & & & \\
&&&&&&\\

Cu$_2$HgSnS$_4$ & 5.61 & 1.95 & 0.83 & -2.78 & 19.10 & 7.08 (6.90, 7.45) \\
& 5.57$^{iii}$\cite{Cu2HgSnS4} & 1.95$^{iii}$\cite{Cu2HgSnS4} & & & & \\
&&&&&&\\

Cu$_2$HgSnSe$_4$ & 5.88 & 1.95 & 0.36 & -2.28 & 18.12 & 9.58 (9.17, 10.40)\\
& 5.83$^{iii}$\cite{130} & 1.96$^{iii}$\cite{130} & 0.17$^{iii}$\cite{149_1} & & & \\
& 5.84$^{iv}$\cite{146} & 1.97$^{iv}$\cite{146} & & & & 13.78$^{iv}$\cite{146} \\
&&&&&&\\

Cu$_2$HgSnTe$_4$ & 6.20 & 1.97 & 0.39 & -1.71 & 16.15 & 11.38 (11.21, 11.70) \\
& 6.19$^{iii}$\cite{130} & 1.98$^{iii}$\cite{130} & & & & \\

\hline

\multicolumn{7}{l}{$^{i}$ The GGA functional is used for the calculation of the formation energy.}\\
\multicolumn{7}{l}{$^{ii}$ $\varepsilon_\infty^{avr.}$ and $\omega_p^{avr.}$ represent the value of an arithmetic average with respect to the direction of polarization.}\\
\multicolumn{7}{l}{$^{iii}$ Experimental results.}\\
\multicolumn{7}{l}{$^{iv}$ Theoretical results using GGA functional.}\\
\multicolumn{7}{l}{$^{v}$ Theoretical results using HSE06 functional.}\\
\multicolumn{7}{l}{$^{vi}$ Theoretical results using GGA functional. A rigid shift is applied to correct
the band gap\cite{Cu2ZnGeTe4_the}}.\\
\multicolumn{7}{l}{$\varepsilon_\infty^{\bot}$, and $\varepsilon_\infty^{\|}$ is the optical dielectric constant along x-, and z- direction respectively.}\\
\multicolumn{7}{l}{(x), and (z) refer to $\varepsilon_\infty$ in the x-, and z-direction, respectively.}\\
\end{longtable*}

%figure-S1*
\begin{figure}[!htp]
\begin{center}
\includegraphics[width=0.9\hsize]{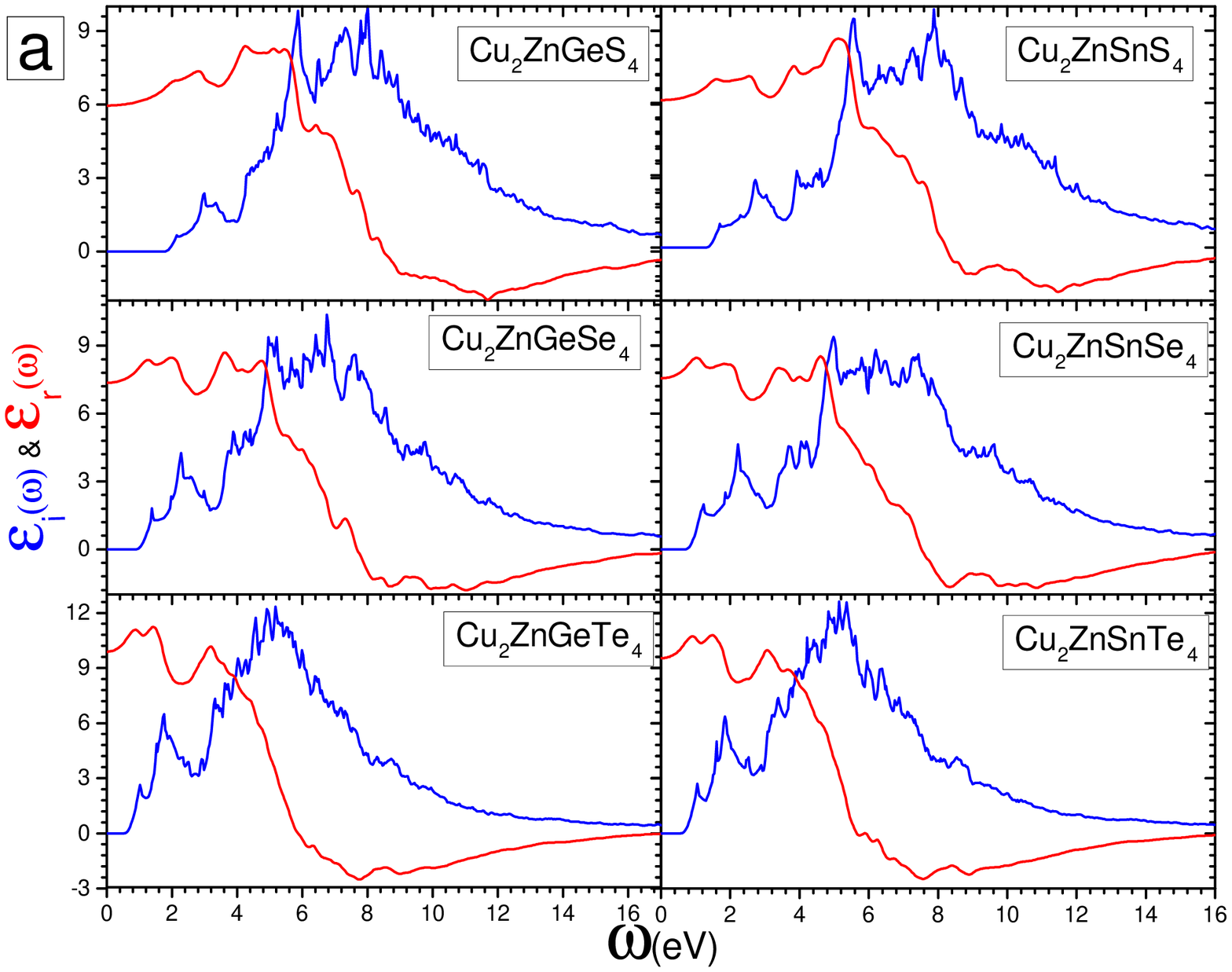}
\includegraphics[width=0.9\hsize]{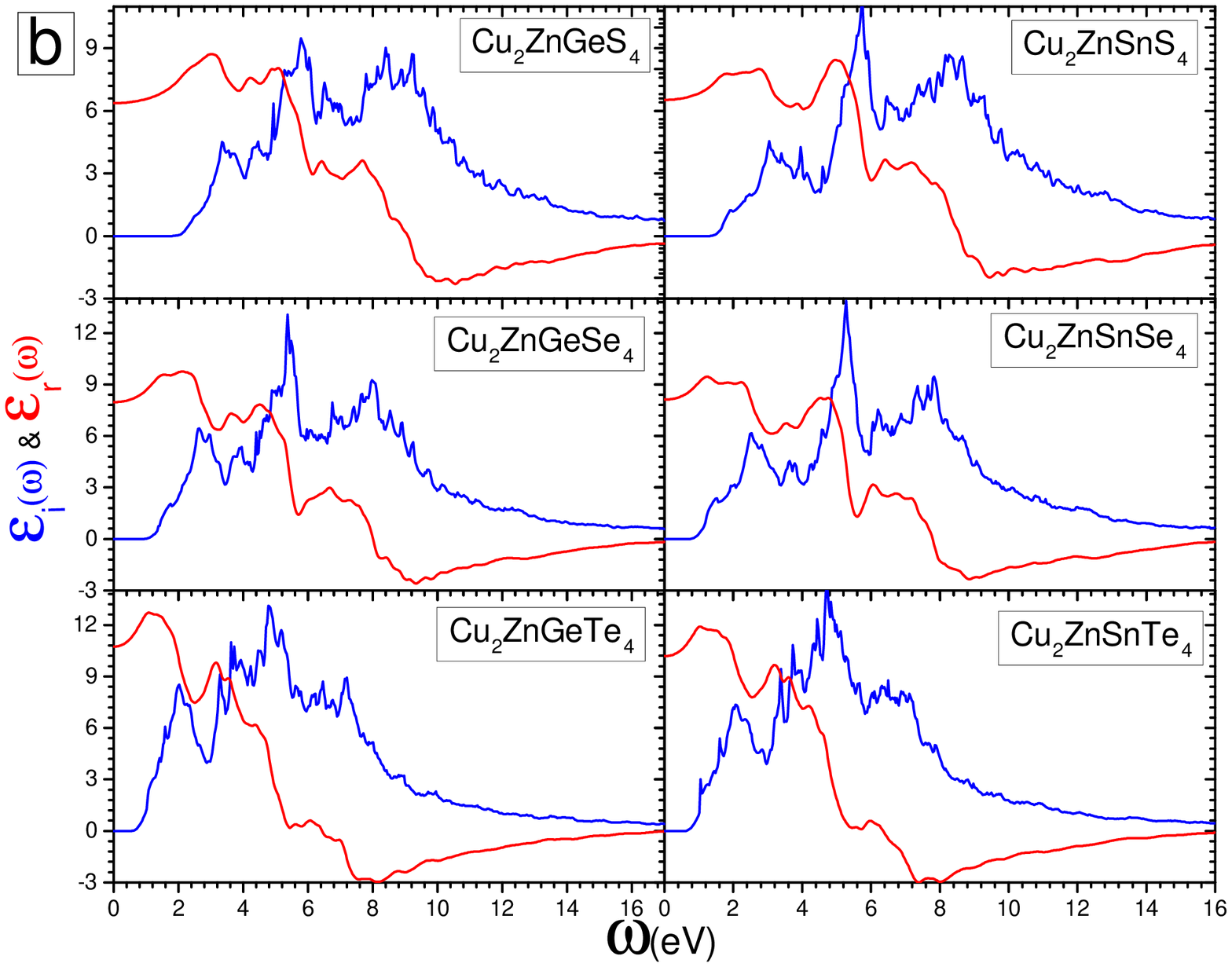}
\caption{\label{fig-S1} Imaginary and real part of the dielectric function
($\varepsilon_i$, and $\varepsilon_r$) along (a) the x-direction, and (b) the z-direction of
the studied Cu$_2$Zn-based chalcogenides. For each compound, the blue, and red figure corresponds
to $\varepsilon_i$, and $\varepsilon_r$, respectively.}
\end{center}
\end{figure}

%figure-S2*
\begin{figure}[!htp]
\begin{center}
\includegraphics[width=0.9\hsize]{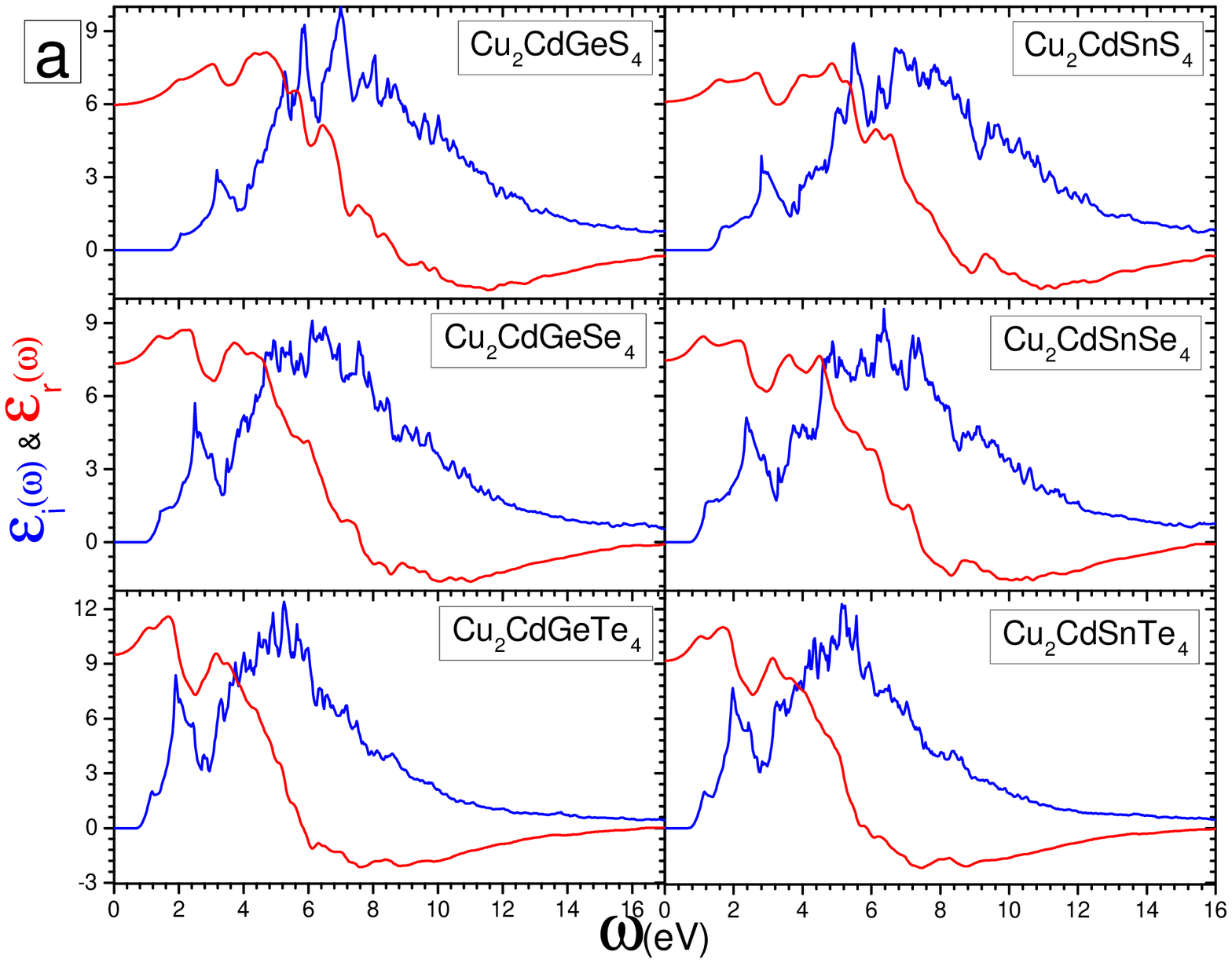}
\includegraphics[width=0.9\hsize]{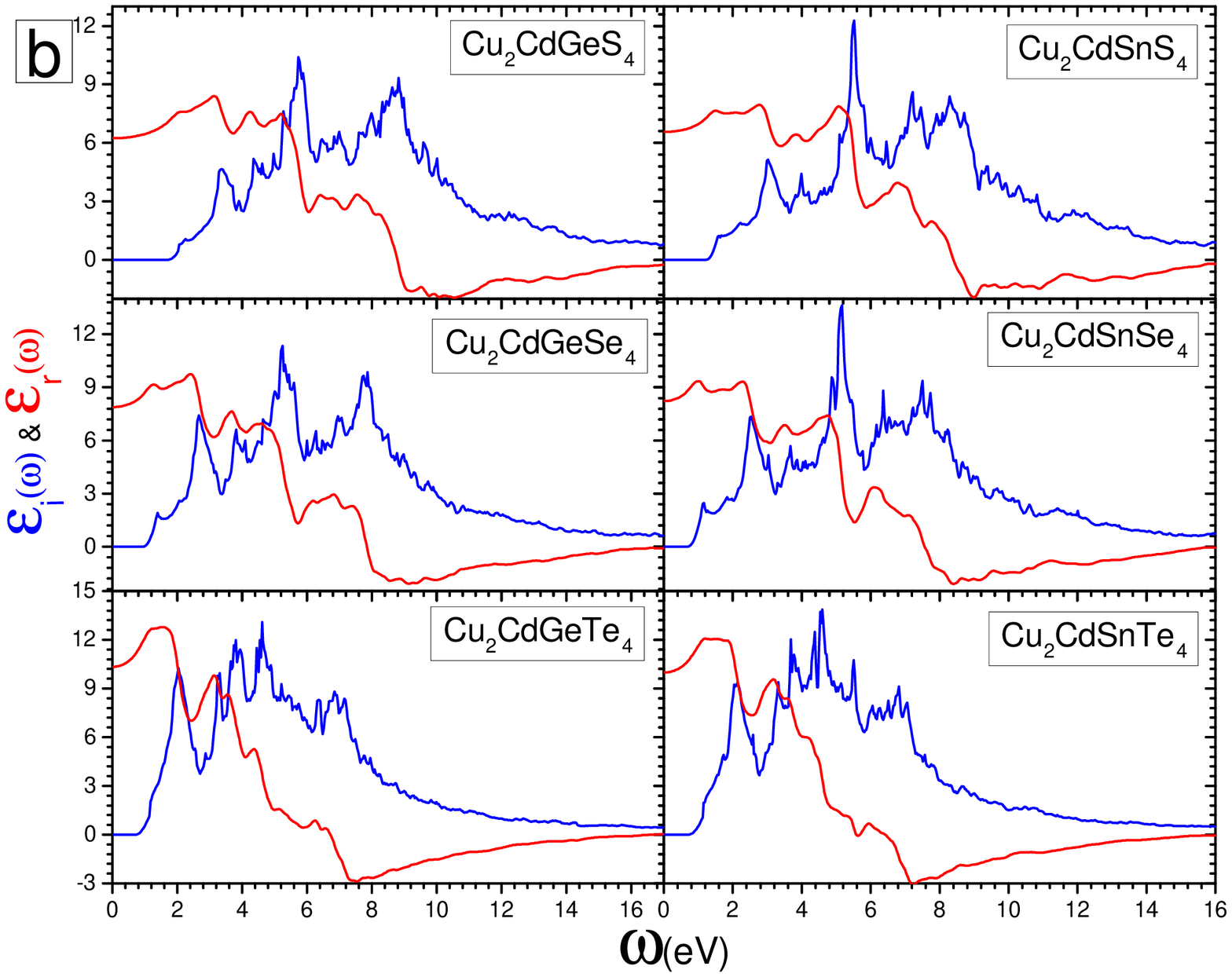}
\caption{Imaginary and real part of the dielectric function
($\varepsilon_i$, and $\varepsilon_r$) along (a) the x-direction, and (b) the z-direction of
the studied Cu$_2$Cd-based chalcogenides. For each compound, the blue, and red figure corresponds
to $\varepsilon_i$, and $\varepsilon_r$, respectively.}
\label{fig-S2}
\end{center}
\end{figure}

%figure-S3*
\begin{figure}[!htp]
\begin{center}
\includegraphics[width=0.9\hsize]{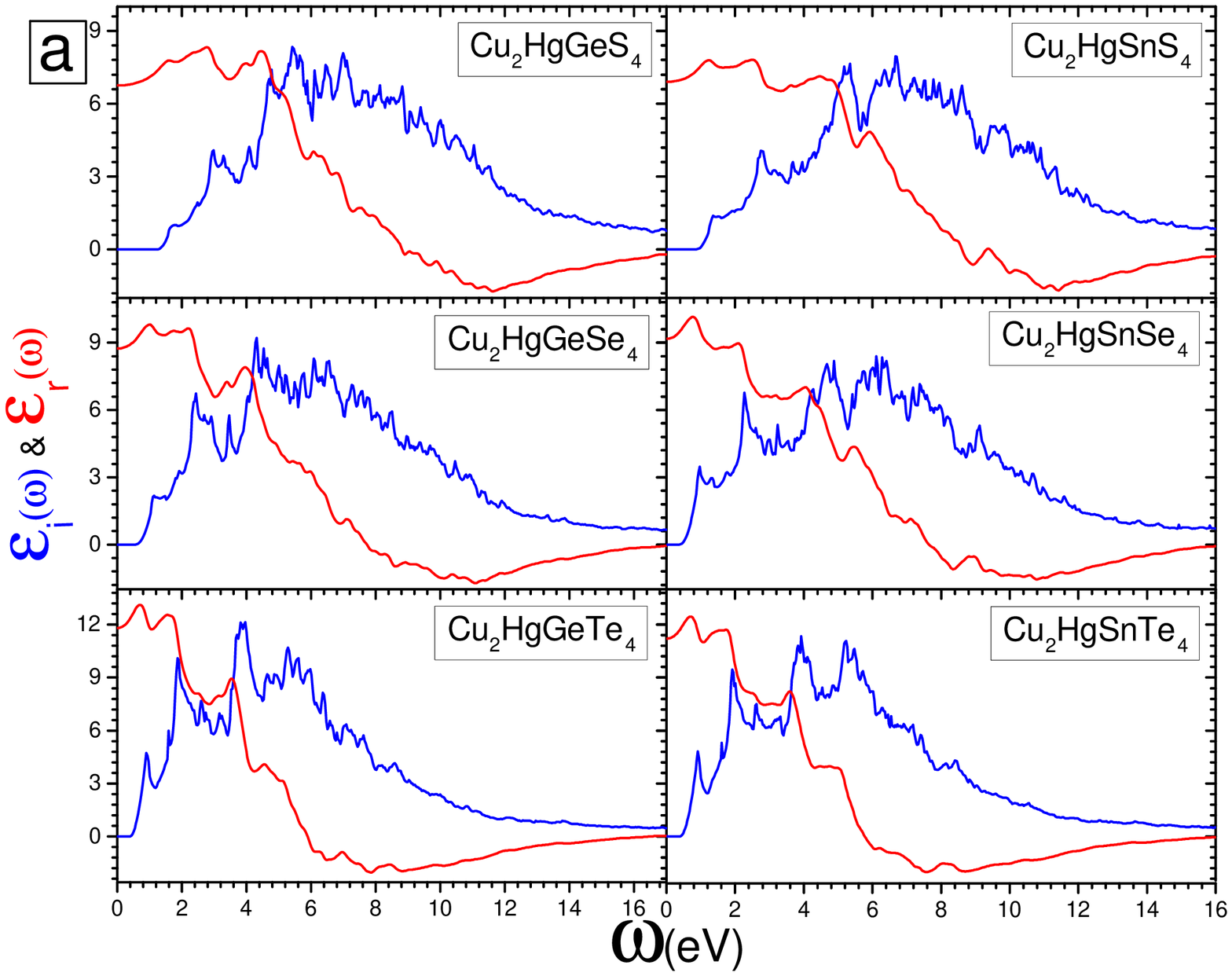}
\includegraphics[width=0.9\hsize]{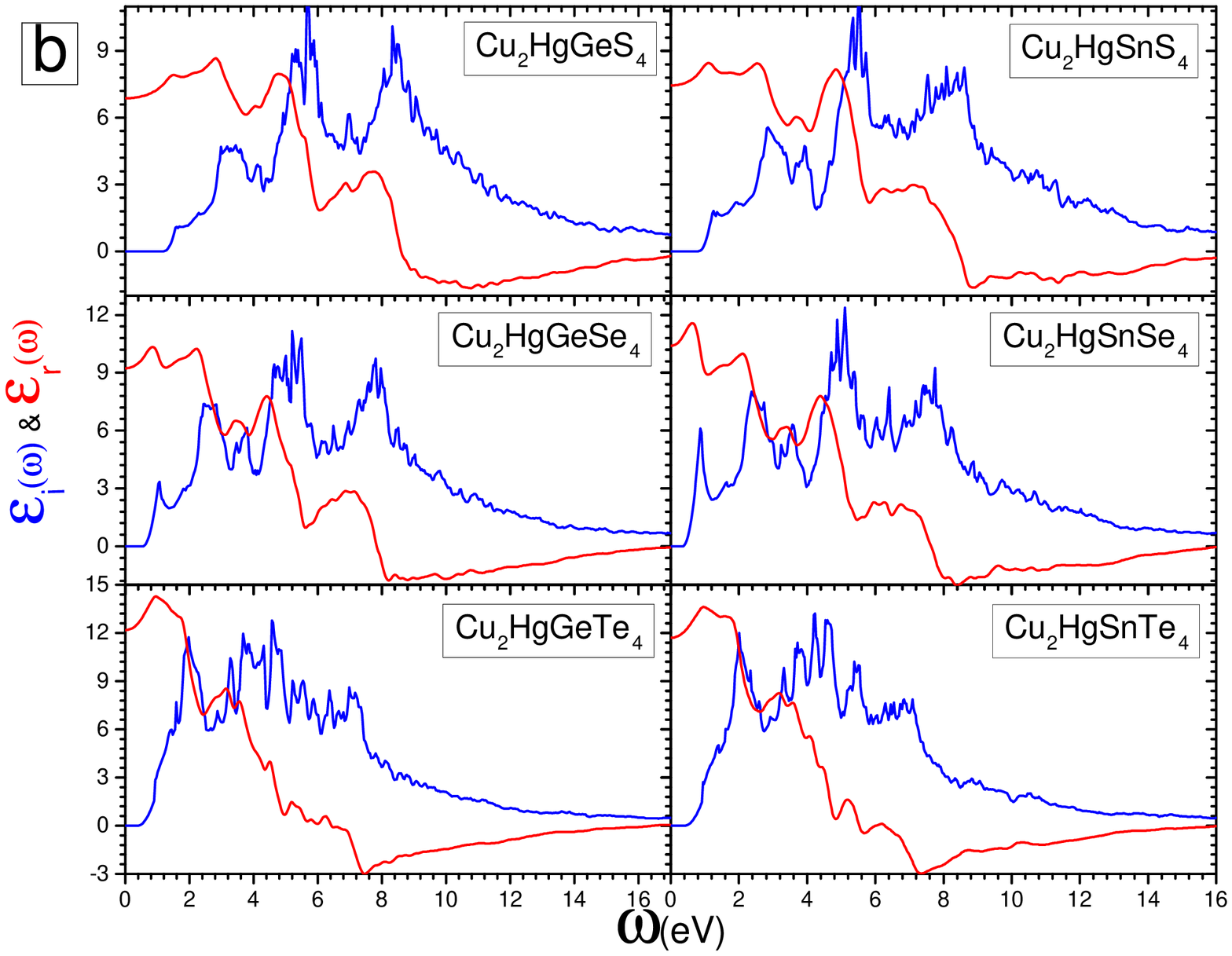}
\caption{\label{fig-S3} Imaginary and real part of the dielectric function
($\varepsilon_i$, and $\varepsilon_r$) along (a) the x-direction, and (b) the z-direction of
the studied Cu$_2$Hg-based chalcogenides. For each compound, the blue, and red figure corresponds
to $\varepsilon_i$,
and $\varepsilon_r$, respectively.}
\end{center}
\end{figure}

%figure-S4*
\begin{figure}[!htp]
\begin{center}
\includegraphics[width=0.9\hsize]{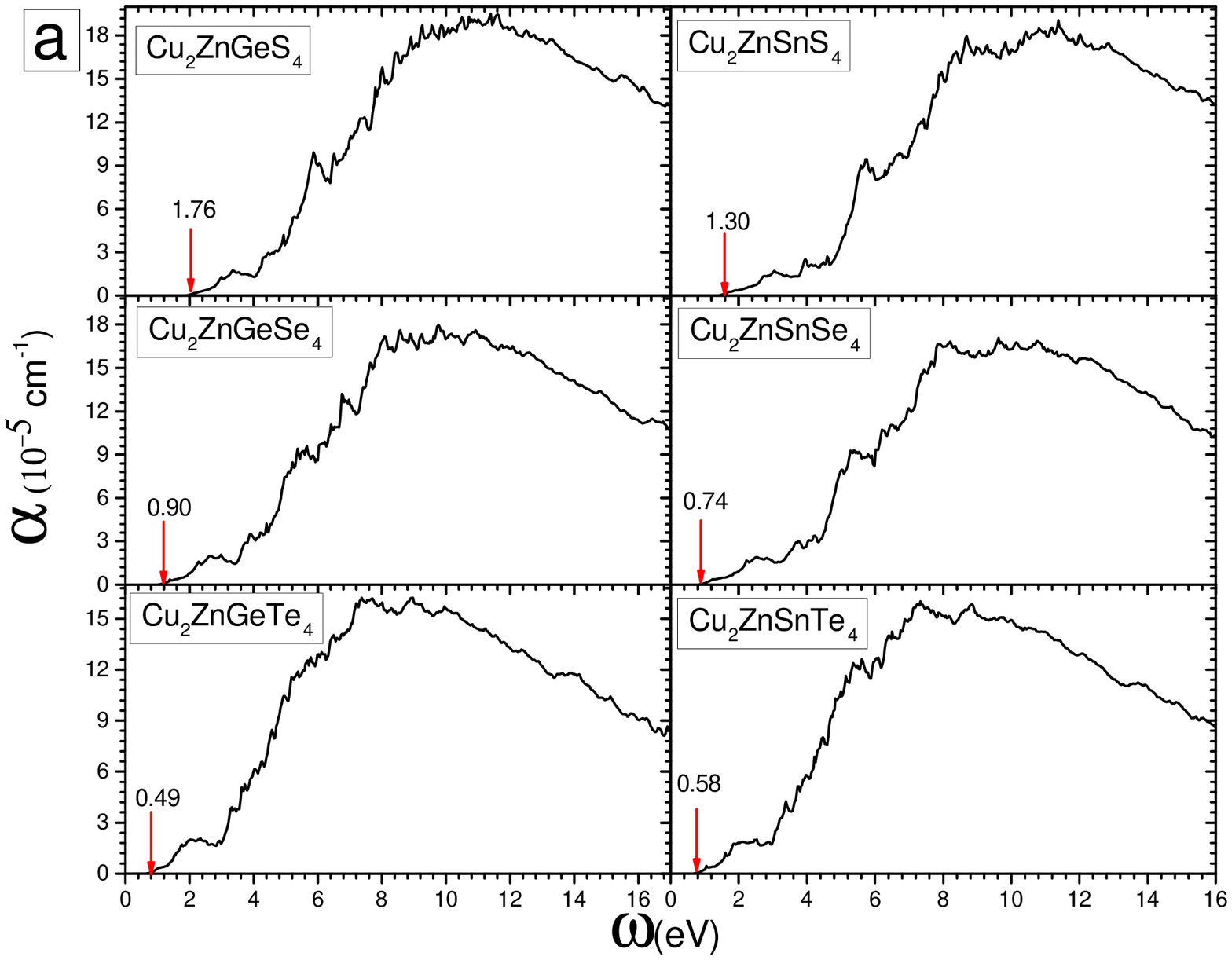}
\includegraphics[width=0.9\hsize]{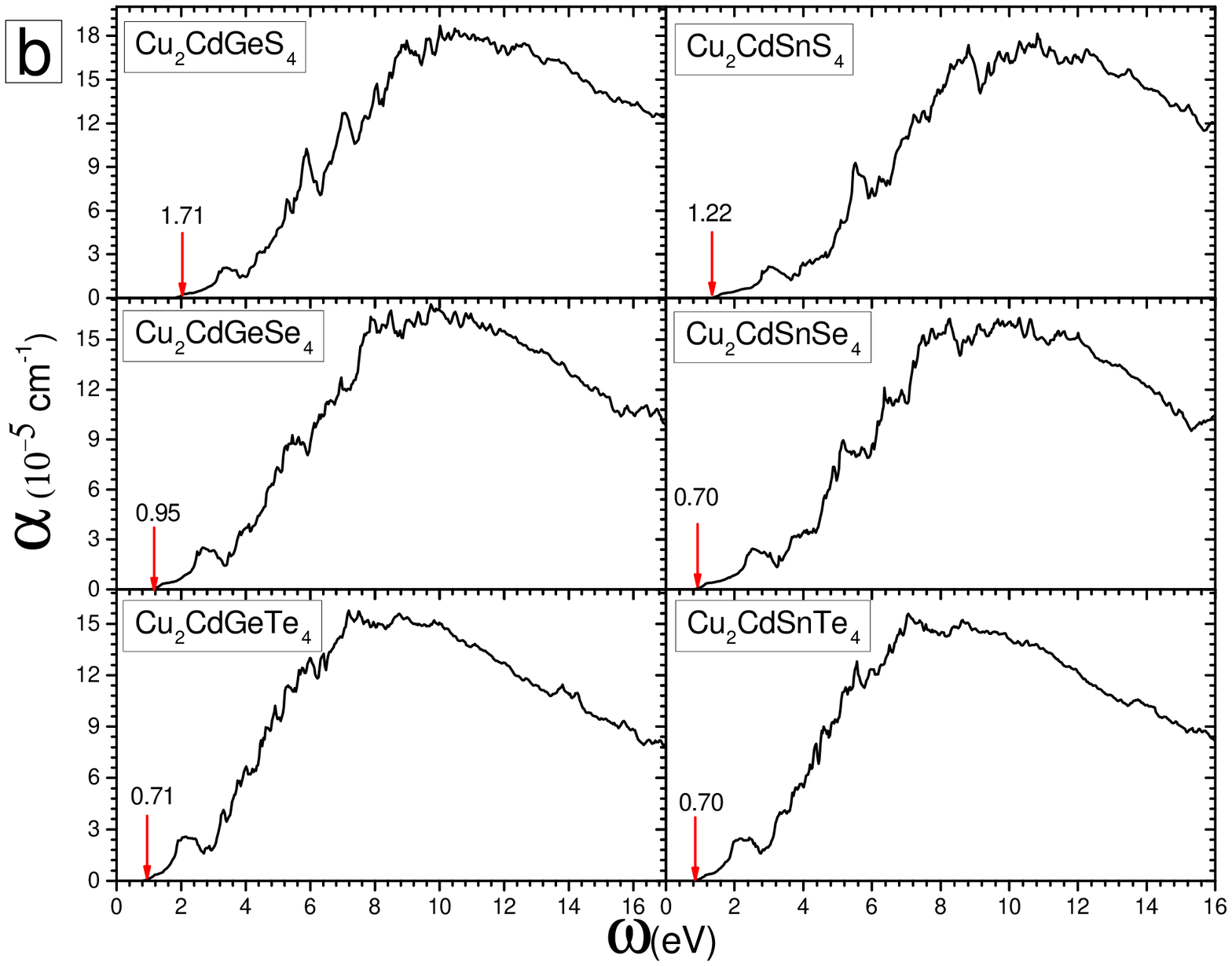}
\includegraphics[width=0.9\hsize]{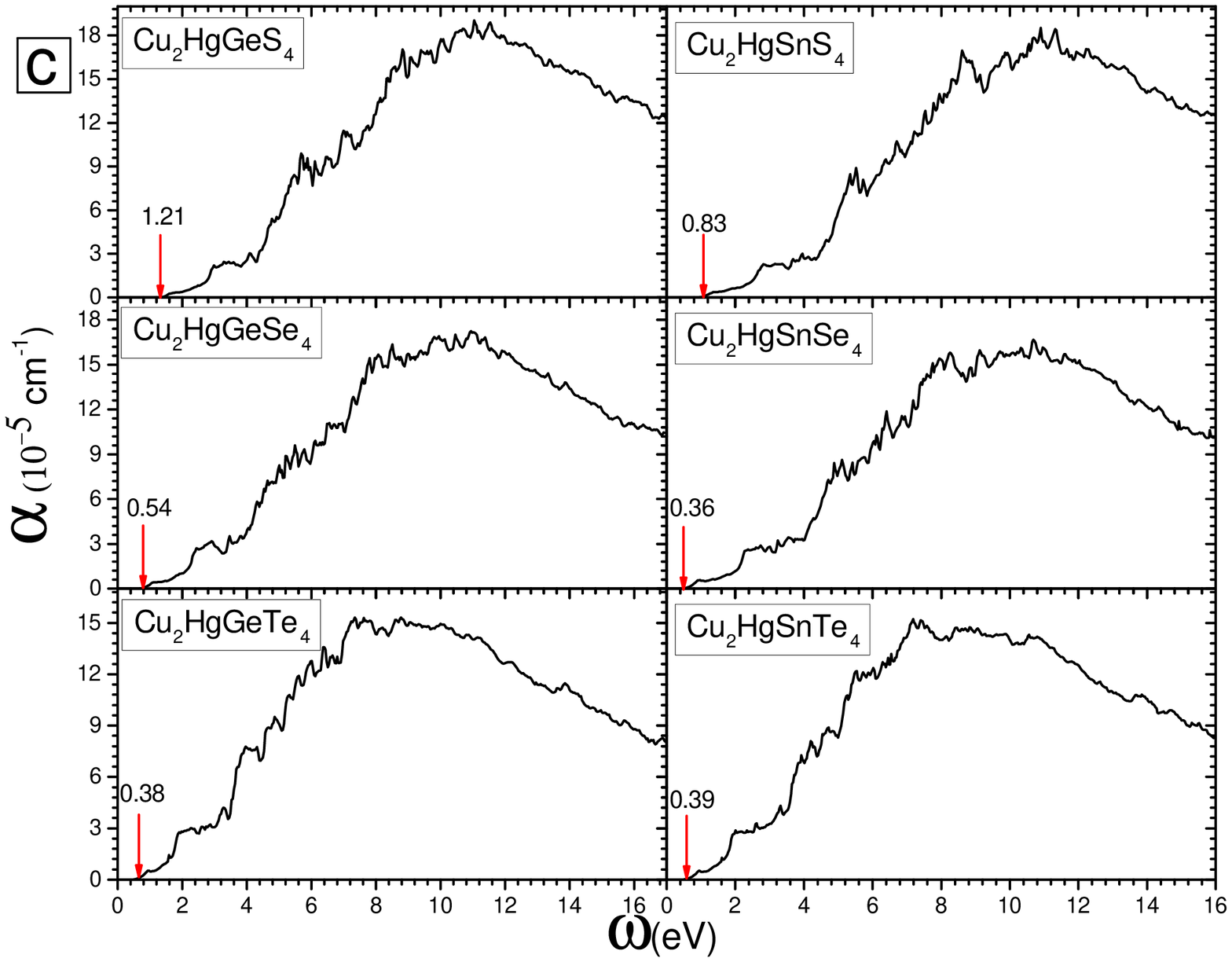}
\caption{\label{fig-S4} Averaged absorption coefficient versus photon energy (eV)
for the studied (a) Cu$_2$Zn-based, (b) Cu$_2$Cd-based, and (a) Cu$_2$Hg-based chalcogenides.
The value of the first direct allowed transition is given for each compound in the corresponding plot.}
\end{center}
\end{figure}

\clearpage

\bibliography{absorber_bib}

%merlin.mbs apsrev4-1.bst 2010-07-25 4.21a (PWD, AO, DPC) hacked
%Control: key (0)
%Control: author (8) initials jnrlst
%Control: editor formatted (1) identically to author
%Control: production of article title (-1) disabled
%Control: page (0) single
%Control: year (1) truncated
%Control: production of eprint (0) enabled
\begin{thebibliography}{58}%
\makeatletter
\providecommand \@ifxundefined [1]{%
 \@ifx{#1\undefined}
}%
\providecommand \@ifnum [1]{%
 \ifnum #1\expandafter \@firstoftwo
 \else \expandafter \@secondoftwo
 \fi
}%
\providecommand \@ifx [1]{%
 \ifx #1\expandafter \@firstoftwo
 \else \expandafter \@secondoftwo
 \fi
}%
\providecommand \natexlab [1]{#1}%
\providecommand \enquote  [1]{``#1''}%
\providecommand \bibnamefont  [1]{#1}%
\providecommand \bibfnamefont [1]{#1}%
\providecommand \citenamefont [1]{#1}%
\providecommand \href@noop [0]{\@secondoftwo}%
\providecommand \href [0]{\begingroup \@sanitize@url \@href}%
\providecommand \@href[1]{\@@startlink{#1}\@@href}%
\providecommand \@@href[1]{\endgroup#1\@@endlink}%
\providecommand \@sanitize@url [0]{\catcode `\\12\catcode `\$12\catcode
  `\&12\catcode `\#12\catcode `\^12\catcode `\_12\catcode `\%12\relax}%
\providecommand \@@startlink[1]{}%
\providecommand \@@endlink[0]{}%
\providecommand \url  [0]{\begingroup\@sanitize@url \@url }%
\providecommand \@url [1]{\endgroup\@href {#1}{\urlprefix }}%
\providecommand \urlprefix  [0]{URL }%
\providecommand \Eprint [0]{\href }%
\providecommand \doibase [0]{http://dx.doi.org/}%
\providecommand \selectlanguage [0]{\@gobble}%
\providecommand \bibinfo  [0]{\@secondoftwo}%
\providecommand \bibfield  [0]{\@secondoftwo}%
\providecommand \translation [1]{[#1]}%
\providecommand \BibitemOpen [0]{}%
\providecommand \bibitemStop [0]{}%
\providecommand \bibitemNoStop [0]{.\EOS\space}%
\providecommand \EOS [0]{\spacefactor3000\relax}%
\providecommand \BibitemShut  [1]{\csname bibitem#1\endcsname}%
\let\auto@bib@innerbib\@empty
%</preamble>
\bibitem [{\citenamefont {Katagiri}\ \emph {et~al.}(2008)\citenamefont
  {Katagiri}, \citenamefont {Jimbo}, \citenamefont {Yamada}, \citenamefont
  {Kamimura}, \citenamefont {Maw}, \citenamefont {Fukano}, \citenamefont
  {Ito},\ and\ \citenamefont {Motohiro}}]{3J}%
  \BibitemOpen
  \bibfield  {author} {\bibinfo {author} {\bibfnamefont {H.}~\bibnamefont
  {Katagiri}}, \bibinfo {author} {\bibfnamefont {K.}~\bibnamefont {Jimbo}},
  \bibinfo {author} {\bibfnamefont {S.}~\bibnamefont {Yamada}}, \bibinfo
  {author} {\bibfnamefont {T.}~\bibnamefont {Kamimura}}, \bibinfo {author}
  {\bibfnamefont {W.~S.}\ \bibnamefont {Maw}}, \bibinfo {author} {\bibfnamefont
  {T.}~\bibnamefont {Fukano}}, \bibinfo {author} {\bibfnamefont
  {T.}~\bibnamefont {Ito}}, \ and\ \bibinfo {author} {\bibfnamefont
  {T.}~\bibnamefont {Motohiro}},\ }\href@noop {} {\bibfield  {journal}
  {\bibinfo  {journal} {Appl. Phys. Express}\ }\textbf {\bibinfo {volume}
  {1}},\ \bibinfo {pages} {041201} (\bibinfo {year} {2008})}\BibitemShut
  {NoStop}%
\bibitem [{\citenamefont {Schorr}\ \emph {et~al.}(2007)\citenamefont {Schorr},
  \citenamefont {Wagner}, \citenamefont {Tovar},\ and\ \citenamefont
  {Sheptyakov}}]{4J}%
  \BibitemOpen
  \bibfield  {author} {\bibinfo {author} {\bibfnamefont {S.}~\bibnamefont
  {Schorr}}, \bibinfo {author} {\bibfnamefont {G.}~\bibnamefont {Wagner}},
  \bibinfo {author} {\bibfnamefont {M.}~\bibnamefont {Tovar}}, \ and\ \bibinfo
  {author} {\bibfnamefont {D.}~\bibnamefont {Sheptyakov}},\ }\href@noop {}
  {\bibfield  {journal} {\bibinfo  {journal} {Mater. Res. Soc. Symp. Proc.}\
  }\textbf {\bibinfo {volume} {1012}},\ \bibinfo {pages} {Y03} (\bibinfo {year}
  {2007})}\BibitemShut {NoStop}%
\bibitem [{\citenamefont {Parasyuk}\ \emph {et~al.}(2005)\citenamefont
  {Parasyuk}, \citenamefont {Olekseyuk},\ and\ \citenamefont
  {Piskach}}]{Cu2ZnGeTe4_exp}%
  \BibitemOpen
  \bibfield  {author} {\bibinfo {author} {\bibfnamefont {O.~V.}\ \bibnamefont
  {Parasyuk}}, \bibinfo {author} {\bibfnamefont {I.~D.}\ \bibnamefont
  {Olekseyuk}}, \ and\ \bibinfo {author} {\bibfnamefont {L.~V.}\ \bibnamefont
  {Piskach}},\ }\href@noop {} {\bibfield  {journal} {\bibinfo  {journal} {J.
  Alloy. Compd.}\ }\textbf {\bibinfo {volume} {397}},\ \bibinfo {pages} {169}
  (\bibinfo {year} {2005})}\BibitemShut {NoStop}%
\bibitem [{\citenamefont {Matsushita}\ \emph {et~al.}(2000)\citenamefont
  {Matsushita}, \citenamefont {Maeda}, \citenamefont {Katsui},\ and\
  \citenamefont {Takizawa}}]{127}%
  \BibitemOpen
  \bibfield  {author} {\bibinfo {author} {\bibfnamefont {H.}~\bibnamefont
  {Matsushita}}, \bibinfo {author} {\bibfnamefont {T.}~\bibnamefont {Maeda}},
  \bibinfo {author} {\bibfnamefont {A.}~\bibnamefont {Katsui}}, \ and\ \bibinfo
  {author} {\bibfnamefont {T.}~\bibnamefont {Takizawa}},\ }\href@noop {}
  {\bibfield  {journal} {\bibinfo  {journal} {J. Cryst. Growth}\ }\textbf
  {\bibinfo {volume} {208}},\ \bibinfo {pages} {416} (\bibinfo {year}
  {2000})}\BibitemShut {NoStop}%
\bibitem [{\citenamefont {Chen}\ \emph {et~al.}(2009)\citenamefont {Chen},
  \citenamefont {Gong}, \citenamefont {Walsh},\ and\ \citenamefont
  {Wei}}]{pv6}%
  \BibitemOpen
  \bibfield  {author} {\bibinfo {author} {\bibfnamefont {S.}~\bibnamefont
  {Chen}}, \bibinfo {author} {\bibfnamefont {X.~G.}\ \bibnamefont {Gong}},
  \bibinfo {author} {\bibfnamefont {A.}~\bibnamefont {Walsh}}, \ and\ \bibinfo
  {author} {\bibfnamefont {S.~H.}\ \bibnamefont {Wei}},\ }\href@noop {}
  {\bibfield  {journal} {\bibinfo  {journal} {Appl. Phys. Lett.}\ }\textbf
  {\bibinfo {volume} {94}},\ \bibinfo {pages} {041903} (\bibinfo {year}
  {2009})}\BibitemShut {NoStop}%
\bibitem [{\citenamefont {Madelung}(2004)}]{12L}%
  \BibitemOpen
  \bibfield  {author} {\bibinfo {author} {\bibfnamefont {O.~M.}\ \bibnamefont
  {Madelung}},\ }\href@noop {} {\emph {\bibinfo {title} {Semiconductors$:$ Data
  Handbook}}},\ \bibinfo {edition} {3rd}\ ed.\ (\bibinfo  {publisher}
  {Springer},\ \bibinfo {address} {New York},\ \bibinfo {year}
  {2004})\BibitemShut {NoStop}%
\bibitem [{\citenamefont {Parasyuk}\ \emph {et~al.}(2001)\citenamefont
  {Parasyuk}, \citenamefont {Gulay}, \citenamefont {Romanyuk},\ and\
  \citenamefont {Piskach}}]{15L}%
  \BibitemOpen
  \bibfield  {author} {\bibinfo {author} {\bibfnamefont {O.~V.}\ \bibnamefont
  {Parasyuk}}, \bibinfo {author} {\bibfnamefont {L.~D.}\ \bibnamefont {Gulay}},
  \bibinfo {author} {\bibfnamefont {Y.~E.}\ \bibnamefont {Romanyuk}}, \ and\
  \bibinfo {author} {\bibfnamefont {L.~V.}\ \bibnamefont {Piskach}},\
  }\href@noop {} {\bibfield  {journal} {\bibinfo  {journal} {J. Alloys Compd.}\
  }\textbf {\bibinfo {volume} {329}},\ \bibinfo {pages} {202} (\bibinfo {year}
  {2001})}\BibitemShut {NoStop}%
\bibitem [{\citenamefont {Todorov}\ \emph {et~al.}(2013)\citenamefont
  {Todorov}, \citenamefont {Tang}, \citenamefont {Bag}, \citenamefont
  {Gunawan}, \citenamefont {Gokmen}, \citenamefont {Zhu},\ and\ \citenamefont
  {Mitzi}}]{2F}%
  \BibitemOpen
  \bibfield  {author} {\bibinfo {author} {\bibfnamefont {T.~K.}\ \bibnamefont
  {Todorov}}, \bibinfo {author} {\bibfnamefont {J.}~\bibnamefont {Tang}},
  \bibinfo {author} {\bibfnamefont {S.}~\bibnamefont {Bag}}, \bibinfo {author}
  {\bibfnamefont {O.}~\bibnamefont {Gunawan}}, \bibinfo {author} {\bibfnamefont
  {T.}~\bibnamefont {Gokmen}}, \bibinfo {author} {\bibfnamefont
  {Y.}~\bibnamefont {Zhu}}, \ and\ \bibinfo {author} {\bibfnamefont {D.~B.}\
  \bibnamefont {Mitzi}},\ }\href@noop {} {\bibfield  {journal} {\bibinfo
  {journal} {Adv. Energy Mater.}\ }\textbf {\bibinfo {volume} {3}},\ \bibinfo
  {pages} {34} (\bibinfo {year} {2013})}\BibitemShut {NoStop}%
\bibitem [{\citenamefont {Barkhouse}\ \emph {et~al.}(2012)\citenamefont
  {Barkhouse}, \citenamefont {Gunawan}, \citenamefont {Gokmen}, \citenamefont
  {Todorov},\ and\ \citenamefont {Mitzi}}]{7G}%
  \BibitemOpen
  \bibfield  {author} {\bibinfo {author} {\bibfnamefont {D.}~\bibnamefont
  {Barkhouse}}, \bibinfo {author} {\bibfnamefont {O.}~\bibnamefont {Gunawan}},
  \bibinfo {author} {\bibfnamefont {T.}~\bibnamefont {Gokmen}}, \bibinfo
  {author} {\bibfnamefont {T.}~\bibnamefont {Todorov}}, \ and\ \bibinfo
  {author} {\bibfnamefont {D.}~\bibnamefont {Mitzi}},\ }\href@noop {}
  {\bibfield  {journal} {\bibinfo  {journal} {Prog. Photovolt. Res. Appl.}\
  }\textbf {\bibinfo {volume} {20}},\ \bibinfo {pages} {6} (\bibinfo {year}
  {2012})}\BibitemShut {NoStop}%
\bibitem [{\citenamefont {Scragg}\ \emph {et~al.}(2008)\citenamefont {Scragg},
  \citenamefont {Dale},\ and\ \citenamefont {Peter}}]{pv5}%
  \BibitemOpen
  \bibfield  {author} {\bibinfo {author} {\bibfnamefont {J.~J.}\ \bibnamefont
  {Scragg}}, \bibinfo {author} {\bibfnamefont {P.~J.}\ \bibnamefont {Dale}}, \
  and\ \bibinfo {author} {\bibfnamefont {L.~M.}\ \bibnamefont {Peter}},\
  }\href@noop {} {\bibfield  {journal} {\bibinfo  {journal} {Electrochem.
  Commun.}\ }\textbf {\bibinfo {volume} {10}},\ \bibinfo {pages} {639}
  (\bibinfo {year} {2008})}\BibitemShut {NoStop}%
\bibitem [{\citenamefont {Altosaar}\ \emph {et~al.}(2008)\citenamefont
  {Altosaar}, \citenamefont {Raudoja}, \citenamefont {Timmo}, \citenamefont
  {Danilson}, \citenamefont {Grossberg}, \citenamefont {Krustok},\ and\
  \citenamefont {Mellikov}}]{22L}%
  \BibitemOpen
  \bibfield  {author} {\bibinfo {author} {\bibfnamefont {M.}~\bibnamefont
  {Altosaar}}, \bibinfo {author} {\bibfnamefont {J.}~\bibnamefont {Raudoja}},
  \bibinfo {author} {\bibfnamefont {K.}~\bibnamefont {Timmo}}, \bibinfo
  {author} {\bibfnamefont {M.}~\bibnamefont {Danilson}}, \bibinfo {author}
  {\bibfnamefont {M.}~\bibnamefont {Grossberg}}, \bibinfo {author}
  {\bibfnamefont {J.}~\bibnamefont {Krustok}}, \ and\ \bibinfo {author}
  {\bibfnamefont {E.}~\bibnamefont {Mellikov}},\ }\href@noop {} {\bibfield
  {journal} {\bibinfo  {journal} {Phys. Status Solidi A}\ }\textbf {\bibinfo
  {volume} {205}},\ \bibinfo {pages} {167} (\bibinfo {year}
  {2008})}\BibitemShut {NoStop}%
\bibitem [{\citenamefont {Siebentritt}(2013)}]{5H}%
  \BibitemOpen
  \bibfield  {author} {\bibinfo {author} {\bibfnamefont {S.}~\bibnamefont
  {Siebentritt}},\ }\href@noop {} {\bibfield  {journal} {\bibinfo  {journal}
  {Thin Solid Films}\ }\textbf {\bibinfo {volume} {535}},\ \bibinfo {pages} {1}
  (\bibinfo {year} {2013})}\BibitemShut {NoStop}%
\bibitem [{\citenamefont {Yu}\ and\ \citenamefont {Zunger}(2012)}]{zunger}%
  \BibitemOpen
  \bibfield  {author} {\bibinfo {author} {\bibfnamefont {L.}~\bibnamefont
  {Yu}}\ and\ \bibinfo {author} {\bibfnamefont {A.}~\bibnamefont {Zunger}},\
  }\href@noop {} {\bibfield  {journal} {\bibinfo  {journal} {Phys. Rev. Lett.}\
  }\textbf {\bibinfo {volume} {108}},\ \bibinfo {pages} {068701} (\bibinfo
  {year} {2012})}\BibitemShut {NoStop}%
\bibitem [{\citenamefont {Yin}\ \emph {et~al.}(2014)\citenamefont {Yin},
  \citenamefont {Shi},\ and\ \citenamefont {Yan}}]{perovskite}%
  \BibitemOpen
  \bibfield  {author} {\bibinfo {author} {\bibfnamefont {W.~J.}\ \bibnamefont
  {Yin}}, \bibinfo {author} {\bibfnamefont {T.}~\bibnamefont {Shi}}, \ and\
  \bibinfo {author} {\bibfnamefont {Y.}~\bibnamefont {Yan}},\ }\href@noop {}
  {\bibfield  {journal} {\bibinfo  {journal} {Adv. Mater.}\ }\textbf {\bibinfo
  {volume} {26}},\ \bibinfo {pages} {4653} (\bibinfo {year}
  {2014})}\BibitemShut {NoStop}%
\bibitem [{\citenamefont {Lee}\ \emph {et~al.}(2014)\citenamefont {Lee},
  \citenamefont {Lee}, \citenamefont {Oh}, \citenamefont {Kim},\ and\
  \citenamefont {Chang}}]{SLME1}%
  \BibitemOpen
  \bibfield  {author} {\bibinfo {author} {\bibfnamefont {I.~H.}\ \bibnamefont
  {Lee}}, \bibinfo {author} {\bibfnamefont {J.}~\bibnamefont {Lee}}, \bibinfo
  {author} {\bibfnamefont {Y.~J.}\ \bibnamefont {Oh}}, \bibinfo {author}
  {\bibfnamefont {S.}~\bibnamefont {Kim}}, \ and\ \bibinfo {author}
  {\bibfnamefont {K.~J.}\ \bibnamefont {Chang}},\ }\href@noop {} {\bibfield
  {journal} {\bibinfo  {journal} {Phys. Rev. B}\ }\textbf {\bibinfo {volume}
  {90}},\ \bibinfo {pages} {115209} (\bibinfo {year} {2014})}\BibitemShut
  {NoStop}%
\bibitem [{\citenamefont {Shockley}\ and\ \citenamefont {Queisser}(1961)}]{SQ}%
  \BibitemOpen
  \bibfield  {author} {\bibinfo {author} {\bibfnamefont {W.}~\bibnamefont
  {Shockley}}\ and\ \bibinfo {author} {\bibfnamefont {H.~J.}\ \bibnamefont
  {Queisser}},\ }\href@noop {} {\bibfield  {journal} {\bibinfo  {journal} {J.
  Appl. Phys.}\ }\textbf {\bibinfo {volume} {32}},\ \bibinfo {pages} {510}
  (\bibinfo {year} {1961})}\BibitemShut {NoStop}%
\bibitem [{\citenamefont {J$\ddot{a}$ger}\ \emph {et~al.}(2015)\citenamefont
  {J$\ddot{a}$ger}, \citenamefont {Romanyuk}, \citenamefont {Bissig},
  \citenamefont {Pianezzi}, \citenamefont {Nishiwaki}, \citenamefont
  {Reinhard}, \citenamefont {Steinhauser}, \citenamefont {Schwenk},\ and\
  \citenamefont {Tiwari}}]{micron_CIGS}%
  \BibitemOpen
  \bibfield  {author} {\bibinfo {author} {\bibfnamefont {T.}~\bibnamefont
  {J$\ddot{a}$ger}}, \bibinfo {author} {\bibfnamefont {Y.~E.}\ \bibnamefont
  {Romanyuk}}, \bibinfo {author} {\bibfnamefont {B.}~\bibnamefont {Bissig}},
  \bibinfo {author} {\bibfnamefont {F.}~\bibnamefont {Pianezzi}}, \bibinfo
  {author} {\bibfnamefont {S.}~\bibnamefont {Nishiwaki}}, \bibinfo {author}
  {\bibfnamefont {P.}~\bibnamefont {Reinhard}}, \bibinfo {author}
  {\bibfnamefont {J.}~\bibnamefont {Steinhauser}}, \bibinfo {author}
  {\bibfnamefont {J.}~\bibnamefont {Schwenk}}, \ and\ \bibinfo {author}
  {\bibfnamefont {A.~N.}\ \bibnamefont {Tiwari}},\ }\href@noop {} {\bibfield
  {journal} {\bibinfo  {journal} {J. App. Phys.}\ }\textbf {\bibinfo {volume}
  {117}},\ \bibinfo {pages} {225303} (\bibinfo {year} {2015})}\BibitemShut
  {NoStop}%
\bibitem [{\citenamefont {Schorr}(2011)}]{13F}%
  \BibitemOpen
  \bibfield  {author} {\bibinfo {author} {\bibfnamefont {S.}~\bibnamefont
  {Schorr}},\ }\href@noop {} {\bibfield  {journal} {\bibinfo  {journal} {Sol.
  Energy Mater. Sol. Cells}\ }\textbf {\bibinfo {volume} {95}},\ \bibinfo
  {pages} {1482} (\bibinfo {year} {2011})}\BibitemShut {NoStop}%
\bibitem [{\citenamefont {Olekseyuk}\ \emph {et~al.}(2002)\citenamefont
  {Olekseyuk}, \citenamefont {Gulay}, \citenamefont {Dydchak}, \citenamefont
  {Piskach}, \citenamefont {Parasyuk},\ and\ \citenamefont {Marchuk}}]{130}%
  \BibitemOpen
  \bibfield  {author} {\bibinfo {author} {\bibfnamefont {I.~D.}\ \bibnamefont
  {Olekseyuk}}, \bibinfo {author} {\bibfnamefont {L.~D.}\ \bibnamefont
  {Gulay}}, \bibinfo {author} {\bibfnamefont {I.~V.}\ \bibnamefont {Dydchak}},
  \bibinfo {author} {\bibfnamefont {L.~V.}\ \bibnamefont {Piskach}}, \bibinfo
  {author} {\bibfnamefont {O.~V.}\ \bibnamefont {Parasyuk}}, \ and\ \bibinfo
  {author} {\bibfnamefont {O.~V.}\ \bibnamefont {Marchuk}},\ }\href@noop {}
  {\bibfield  {journal} {\bibinfo  {journal} {J. Allo. Comp.}\ }\textbf
  {\bibinfo {volume} {340}},\ \bibinfo {pages} {141} (\bibinfo {year}
  {2002})}\BibitemShut {NoStop}%
\bibitem [{\citenamefont {Liu}\ \emph {et~al.}(2009{\natexlab{a}})\citenamefont
  {Liu}, \citenamefont {Chen}, \citenamefont {Huang},\ and\ \citenamefont
  {Chen}}]{145}%
  \BibitemOpen
  \bibfield  {author} {\bibinfo {author} {\bibfnamefont {M.~L.}\ \bibnamefont
  {Liu}}, \bibinfo {author} {\bibfnamefont {I.~W.}\ \bibnamefont {Chen}},
  \bibinfo {author} {\bibfnamefont {F.~Q.}\ \bibnamefont {Huang}}, \ and\
  \bibinfo {author} {\bibfnamefont {L.~D.}\ \bibnamefont {Chen}},\ }\href@noop
  {} {\bibfield  {journal} {\bibinfo  {journal} {Adv. Mater.}\ }\textbf
  {\bibinfo {volume} {21}},\ \bibinfo {pages} {3808} (\bibinfo {year}
  {2009}{\natexlab{a}})}\BibitemShut {NoStop}%
\bibitem [{\citenamefont {Hohenberg}\ and\ \citenamefont
  {Kohn}(1964)}]{Hohenberg_Kohn1}%
  \BibitemOpen
  \bibfield  {author} {\bibinfo {author} {\bibfnamefont {P.}~\bibnamefont
  {Hohenberg}}\ and\ \bibinfo {author} {\bibfnamefont {W.}~\bibnamefont
  {Kohn}},\ }\href@noop {} {\bibfield  {journal} {\bibinfo  {journal} {Phys.
  Rev.}\ }\textbf {\bibinfo {volume} {136}},\ \bibinfo {pages} {B864} (\bibinfo
  {year} {1964})}\BibitemShut {NoStop}%
\bibitem [{\citenamefont {Kohn}\ and\ \citenamefont
  {Sham}(1965)}]{kohn_sham_1965}%
  \BibitemOpen
  \bibfield  {author} {\bibinfo {author} {\bibfnamefont {W.}~\bibnamefont
  {Kohn}}\ and\ \bibinfo {author} {\bibfnamefont {L.~J.}\ \bibnamefont
  {Sham}},\ }\href@noop {} {\bibfield  {journal} {\bibinfo  {journal} {Phys.
  Rev.}\ }\textbf {\bibinfo {volume} {140}},\ \bibinfo {pages} {A1133}
  (\bibinfo {year} {1965})}\BibitemShut {NoStop}%
\bibitem [{\citenamefont {Kresse}\ and\ \citenamefont {Hafner}(1993)}]{c3}%
  \BibitemOpen
  \bibfield  {author} {\bibinfo {author} {\bibfnamefont {G.}~\bibnamefont
  {Kresse}}\ and\ \bibinfo {author} {\bibfnamefont {J.}~\bibnamefont
  {Hafner}},\ }\href@noop {} {\bibfield  {journal} {\bibinfo  {journal} {Phys.
  Rev. B}\ }\textbf {\bibinfo {volume} {47}},\ \bibinfo {pages} {R558}
  (\bibinfo {year} {1993})}\BibitemShut {NoStop}%
\bibitem [{\citenamefont {Kresse}\ and\ \citenamefont {Hafner}(1994)}]{10b}%
  \BibitemOpen
  \bibfield  {author} {\bibinfo {author} {\bibfnamefont {G.}~\bibnamefont
  {Kresse}}\ and\ \bibinfo {author} {\bibfnamefont {J.}~\bibnamefont
  {Hafner}},\ }\href@noop {} {\bibfield  {journal} {\bibinfo  {journal} {J.
  Phys. Cond. Matt.}\ }\textbf {\bibinfo {volume} {6}},\ \bibinfo {pages}
  {8245} (\bibinfo {year} {1994})}\BibitemShut {NoStop}%
\bibitem [{\citenamefont {Kresse}\ and\ \citenamefont
  {Furthm\"{u}ller}(1996{\natexlab{a}})}]{vasp}%
  \BibitemOpen
  \bibfield  {author} {\bibinfo {author} {\bibfnamefont {G.}~\bibnamefont
  {Kresse}}\ and\ \bibinfo {author} {\bibfnamefont {J.}~\bibnamefont
  {Furthm\"{u}ller}},\ }\href@noop {} {\bibfield  {journal} {\bibinfo
  {journal} {Comput. Mater. Sci.}\ }\textbf {\bibinfo {volume} {6}},\ \bibinfo
  {pages} {15} (\bibinfo {year} {1996}{\natexlab{a}})}\BibitemShut {NoStop}%
\bibitem [{\citenamefont {Kresse}\ and\ \citenamefont
  {Furthm\"{u}ller}(1996{\natexlab{b}})}]{ch2_78}%
  \BibitemOpen
  \bibfield  {author} {\bibinfo {author} {\bibfnamefont {G.}~\bibnamefont
  {Kresse}}\ and\ \bibinfo {author} {\bibfnamefont {J.}~\bibnamefont
  {Furthm\"{u}ller}},\ }\href@noop {} {\bibfield  {journal} {\bibinfo
  {journal} {Phys. Rev. B}\ }\textbf {\bibinfo {volume} {54}},\ \bibinfo
  {pages} {11169} (\bibinfo {year} {1996}{\natexlab{b}})}\BibitemShut {NoStop}%
\bibitem [{\citenamefont {Kresse}\ and\ \citenamefont {Joubert}(1999)}]{c6}%
  \BibitemOpen
  \bibfield  {author} {\bibinfo {author} {\bibfnamefont {G.}~\bibnamefont
  {Kresse}}\ and\ \bibinfo {author} {\bibfnamefont {D.}~\bibnamefont
  {Joubert}},\ }\href@noop {} {\bibfield  {journal} {\bibinfo  {journal} {Phys.
  Rev. B}\ }\textbf {\bibinfo {volume} {59}},\ \bibinfo {pages} {1758}
  (\bibinfo {year} {1999})}\BibitemShut {NoStop}%
\bibitem [{\citenamefont {Bl\"{o}chl}(1994)}]{PAW}%
  \BibitemOpen
  \bibfield  {author} {\bibinfo {author} {\bibfnamefont {P.~E.}\ \bibnamefont
  {Bl\"{o}chl}},\ }\href@noop {} {\bibfield  {journal} {\bibinfo  {journal}
  {Phys. Rev. B}\ }\textbf {\bibinfo {volume} {50}},\ \bibinfo {pages} {17953}
  (\bibinfo {year} {1994})}\BibitemShut {NoStop}%
\bibitem [{\citenamefont {Perdew}\ \emph {et~al.}(1996)\citenamefont {Perdew},
  \citenamefont {Burke},\ and\ \citenamefont {Ernzerhof}}]{PBE_GGA}%
  \BibitemOpen
  \bibfield  {author} {\bibinfo {author} {\bibfnamefont {J.~P.}\ \bibnamefont
  {Perdew}}, \bibinfo {author} {\bibfnamefont {K.}~\bibnamefont {Burke}}, \
  and\ \bibinfo {author} {\bibfnamefont {M.}~\bibnamefont {Ernzerhof}},\
  }\href@noop {} {\bibfield  {journal} {\bibinfo  {journal} {Phys. Rev. Lett.}\
  }\textbf {\bibinfo {volume} {77}},\ \bibinfo {pages} {3865} (\bibinfo {year}
  {1996})}\BibitemShut {NoStop}%
\bibitem [{\citenamefont {Wang}\ and\ \citenamefont {Pickett}(1983)}]{19}%
  \BibitemOpen
  \bibfield  {author} {\bibinfo {author} {\bibfnamefont {C.~S.}\ \bibnamefont
  {Wang}}\ and\ \bibinfo {author} {\bibfnamefont {W.~E.}\ \bibnamefont
  {Pickett}},\ }\href@noop {} {\bibfield  {journal} {\bibinfo  {journal} {Phys.
  Rev. Lett.}\ }\textbf {\bibinfo {volume} {51}},\ \bibinfo {pages} {597}
  (\bibinfo {year} {1983})}\BibitemShut {NoStop}%
\bibitem [{\citenamefont {Monkhorst}\ and\ \citenamefont {Pack}(1976)}]{MP}%
  \BibitemOpen
  \bibfield  {author} {\bibinfo {author} {\bibfnamefont {H.~J.}\ \bibnamefont
  {Monkhorst}}\ and\ \bibinfo {author} {\bibfnamefont {J.~D.}\ \bibnamefont
  {Pack}},\ }\href@noop {} {\bibfield  {journal} {\bibinfo  {journal} {Phys.
  Rev. B}\ }\textbf {\bibinfo {volume} {13}},\ \bibinfo {pages} {5188}
  (\bibinfo {year} {1976})}\BibitemShut {NoStop}%
\bibitem [{\citenamefont {Gajdos}\ \emph {et~al.}(2006)\citenamefont {Gajdos},
  \citenamefont {Hummer}, \citenamefont {Kresse}, \citenamefont
  {Furthm\"{u}ller},\ and\ \citenamefont {Bechstedt}}]{c9}%
  \BibitemOpen
  \bibfield  {author} {\bibinfo {author} {\bibfnamefont {M.}~\bibnamefont
  {Gajdos}}, \bibinfo {author} {\bibfnamefont {K.}~\bibnamefont {Hummer}},
  \bibinfo {author} {\bibfnamefont {G.}~\bibnamefont {Kresse}}, \bibinfo
  {author} {\bibfnamefont {J.}~\bibnamefont {Furthm\"{u}ller}}, \ and\ \bibinfo
  {author} {\bibfnamefont {F.}~\bibnamefont {Bechstedt}},\ }\href@noop {}
  {\bibfield  {journal} {\bibinfo  {journal} {Phys. Rev. B}\ }\textbf {\bibinfo
  {volume} {73}},\ \bibinfo {pages} {045112} (\bibinfo {year}
  {2006})}\BibitemShut {NoStop}%
\bibitem [{\citenamefont {Johnson}\ and\ \citenamefont {Ashcroft}(1998)}]{18}%
  \BibitemOpen
  \bibfield  {author} {\bibinfo {author} {\bibfnamefont {K.~A.}\ \bibnamefont
  {Johnson}}\ and\ \bibinfo {author} {\bibfnamefont {N.~W.}\ \bibnamefont
  {Ashcroft}},\ }\href@noop {} {\bibfield  {journal} {\bibinfo  {journal}
  {Phys. Rev. B}\ }\textbf {\bibinfo {volume} {58}},\ \bibinfo {pages} {15548}
  (\bibinfo {year} {1998})}\BibitemShut {NoStop}%
\bibitem [{\citenamefont {Henderson}\ \emph {et~al.}(2011)\citenamefont
  {Henderson}, \citenamefont {Paier},\ and\ \citenamefont {Scuseria}}]{20}%
  \BibitemOpen
  \bibfield  {author} {\bibinfo {author} {\bibfnamefont {T.~M.}\ \bibnamefont
  {Henderson}}, \bibinfo {author} {\bibfnamefont {J.}~\bibnamefont {Paier}}, \
  and\ \bibinfo {author} {\bibfnamefont {G.~E.}\ \bibnamefont {Scuseria}},\
  }\href@noop {} {\bibfield  {journal} {\bibinfo  {journal} {Phys. Status
  Solidi B}\ }\textbf {\bibinfo {volume} {248}},\ \bibinfo {pages} {767}
  (\bibinfo {year} {2011})}\BibitemShut {NoStop}%
\bibitem [{\citenamefont {Tang}\ \emph {et~al.}(2012)\citenamefont {Tang},
  \citenamefont {Zhu}, \citenamefont {Xue}, \citenamefont {Lu}, \citenamefont
  {Feng}, \citenamefont {Wang},\ and\ \citenamefont
  {Wang}}]{dielectric_exp_PBE_HSE1}%
  \BibitemOpen
  \bibfield  {author} {\bibinfo {author} {\bibfnamefont {F.~L.}\ \bibnamefont
  {Tang}}, \bibinfo {author} {\bibfnamefont {Z.~X.}\ \bibnamefont {Zhu}},
  \bibinfo {author} {\bibfnamefont {H.~T.}\ \bibnamefont {Xue}}, \bibinfo
  {author} {\bibfnamefont {W.~J.}\ \bibnamefont {Lu}}, \bibinfo {author}
  {\bibfnamefont {Y.~D.}\ \bibnamefont {Feng}}, \bibinfo {author}
  {\bibfnamefont {Z.~M.}\ \bibnamefont {Wang}}, \ and\ \bibinfo {author}
  {\bibfnamefont {Y.}~\bibnamefont {Wang}},\ }\href@noop {} {\bibfield
  {journal} {\bibinfo  {journal} {Physica B}\ }\textbf {\bibinfo {volume}
  {407}},\ \bibinfo {pages} {4814} (\bibinfo {year} {2012})}\BibitemShut
  {NoStop}%
\bibitem [{\citenamefont {Wan}\ \emph {et~al.}(2013)\citenamefont {Wan},
  \citenamefont {Tang}, \citenamefont {Zhu}, \citenamefont {Xue}, \citenamefont
  {Lu}, \citenamefont {Feng},\ and\ \citenamefont
  {Rui}}]{dielectric_exp_PBE_HSE2}%
  \BibitemOpen
  \bibfield  {author} {\bibinfo {author} {\bibfnamefont {F.~C.}\ \bibnamefont
  {Wan}}, \bibinfo {author} {\bibfnamefont {F.~L.}\ \bibnamefont {Tang}},
  \bibinfo {author} {\bibfnamefont {Z.~X.}\ \bibnamefont {Zhu}}, \bibinfo
  {author} {\bibfnamefont {H.~T.}\ \bibnamefont {Xue}}, \bibinfo {author}
  {\bibfnamefont {W.~J.}\ \bibnamefont {Lu}}, \bibinfo {author} {\bibfnamefont
  {Y.~D.}\ \bibnamefont {Feng}}, \ and\ \bibinfo {author} {\bibfnamefont
  {Z.~Y.}\ \bibnamefont {Rui}},\ }\href@noop {} {\bibfield  {journal} {\bibinfo
   {journal} {Mat. Sci. Semicon. Proc.}\ }\textbf {\bibinfo {volume} {16}},\
  \bibinfo {pages} {1422} (\bibinfo {year} {2013})}\BibitemShut {NoStop}%
\bibitem [{\citenamefont {Green}(2003)}]{efficiency1}%
  \BibitemOpen
  \bibfield  {author} {\bibinfo {author} {\bibfnamefont {M.~A.}\ \bibnamefont
  {Green}},\ }\href@noop {} {\emph {\bibinfo {title} {Third Generation
  Photovoltaics: advanced solar energy conversion}}}\ (\bibinfo  {publisher}
  {Springer},\ \bibinfo {address} {New York},\ \bibinfo {year}
  {2003})\BibitemShut {NoStop}%
\bibitem [{AM1(2013)}]{AM1.5}%
  \BibitemOpen
  \href@noop {} {\enquote {\bibinfo {title} {{R}eference {S}olar {S}pectral
  {I}rradiance: {A}ir {M}ass 1.5},}\ }\bibinfo {howpublished}
  {\url{http://rredc.nrel.gov/solar/spectra/am1.5/}} (\bibinfo {year} {Date of
  access: 02/08/2013})\BibitemShut {NoStop}%
\bibitem [{\citenamefont {Tiedje}\ \emph {et~al.}(1984)\citenamefont {Tiedje},
  \citenamefont {Yablonovitch}, \citenamefont {Cody},\ and\ \citenamefont
  {Brooks}}]{efficiency2}%
  \BibitemOpen
  \bibfield  {author} {\bibinfo {author} {\bibfnamefont {T.}~\bibnamefont
  {Tiedje}}, \bibinfo {author} {\bibfnamefont {E.}~\bibnamefont
  {Yablonovitch}}, \bibinfo {author} {\bibfnamefont {G.}~\bibnamefont {Cody}},
  \ and\ \bibinfo {author} {\bibfnamefont {B.}~\bibnamefont {Brooks}},\
  }\href@noop {} {\bibfield  {journal} {\bibinfo  {journal} {IEEE Trans.
  Electron}\ }\textbf {\bibinfo {volume} {ED-31}},\ \bibinfo {pages} {711}
  (\bibinfo {year} {1984})}\BibitemShut {NoStop}%
\bibitem [{\citenamefont {Bercx}\ \emph {et~al.}()\citenamefont {Bercx},
  \citenamefont {Sarmadian}, \citenamefont {Saniz}, \citenamefont {Partoens},\
  and\ \citenamefont {Lamoen}}]{Marnik-CuAu}%
  \BibitemOpen
  \bibfield  {author} {\bibinfo {author} {\bibfnamefont {M.}~\bibnamefont
  {Bercx}}, \bibinfo {author} {\bibfnamefont {N.}~\bibnamefont {Sarmadian}},
  \bibinfo {author} {\bibfnamefont {R.}~\bibnamefont {Saniz}}, \bibinfo
  {author} {\bibfnamefont {B.}~\bibnamefont {Partoens}}, \ and\ \bibinfo
  {author} {\bibfnamefont {D.}~\bibnamefont {Lamoen}},\ }\href@noop {}
  {\bibinfo  {journal} {submitted for the publication}\ }\BibitemShut {NoStop}%
\bibitem [{\citenamefont {Wang}\ \emph {et~al.}(2013)\citenamefont {Wang},
  \citenamefont {Winkler}, \citenamefont {Gunawan}, \citenamefont {Gokmen},
  \citenamefont {Todorov}, \citenamefont {Zhu},\ and\ \citenamefont
  {Mitzi}}]{CZTS_12.6}%
  \BibitemOpen
\bibfield  {journal} {  }\bibfield  {author} {\bibinfo {author} {\bibfnamefont
  {W.}~\bibnamefont {Wang}}, \bibinfo {author} {\bibfnamefont {M.~T.}\
  \bibnamefont {Winkler}}, \bibinfo {author} {\bibfnamefont {O.}~\bibnamefont
  {Gunawan}}, \bibinfo {author} {\bibfnamefont {T.}~\bibnamefont {Gokmen}},
  \bibinfo {author} {\bibfnamefont {T.~K.}\ \bibnamefont {Todorov}}, \bibinfo
  {author} {\bibfnamefont {Y.}~\bibnamefont {Zhu}}, \ and\ \bibinfo {author}
  {\bibfnamefont {D.~B.}\ \bibnamefont {Mitzi}},\ }\href@noop {} {\bibfield
  {journal} {\bibinfo  {journal} {Adv. Energy Mater.}\ }\textbf {\bibinfo
  {volume} {4}},\ \bibinfo {pages} {1301465} (\bibinfo {year}
  {2013})}\BibitemShut {NoStop}%
\bibitem [{\citenamefont {Chen}\ and\ \citenamefont
  {Ravindra}(2013)}]{Cu2ZnGeTe4_the}%
  \BibitemOpen
  \bibfield  {author} {\bibinfo {author} {\bibfnamefont {D.}~\bibnamefont
  {Chen}}\ and\ \bibinfo {author} {\bibfnamefont {N.~M.}\ \bibnamefont
  {Ravindra}},\ }\href@noop {} {\bibfield  {journal} {\bibinfo  {journal} {J.
  Alloy. Compd.}\ }\textbf {\bibinfo {volume} {579}},\ \bibinfo {pages} {468}
  (\bibinfo {year} {2013})}\BibitemShut {NoStop}%
\bibitem [{\citenamefont {Bhaskar}\ \emph {et~al.}(2013)\citenamefont
  {Bhaskar}, \citenamefont {Babu}, \citenamefont {Kumar},\ and\ \citenamefont
  {Raja}}]{117}%
  \BibitemOpen
  \bibfield  {author} {\bibinfo {author} {\bibfnamefont {P.~U.}\ \bibnamefont
  {Bhaskar}}, \bibinfo {author} {\bibfnamefont {G.~S.}\ \bibnamefont {Babu}},
  \bibinfo {author} {\bibfnamefont {Y.~B.~K.}\ \bibnamefont {Kumar}}, \ and\
  \bibinfo {author} {\bibfnamefont {V.~S.}\ \bibnamefont {Raja}},\ }\href@noop
  {} {\bibfield  {journal} {\bibinfo  {journal} {Thin Solid Films}\ }\textbf
  {\bibinfo {volume} {534}},\ \bibinfo {pages} {249} (\bibinfo {year}
  {2013})}\BibitemShut {NoStop}%
\bibitem [{\citenamefont {Paier}\ \emph {et~al.}(2009)\citenamefont {Paier},
  \citenamefont {Asahi}, \citenamefont {Nagoya},\ and\ \citenamefont
  {Kresse}}]{pv8}%
  \BibitemOpen
  \bibfield  {author} {\bibinfo {author} {\bibfnamefont {J.}~\bibnamefont
  {Paier}}, \bibinfo {author} {\bibfnamefont {R.}~\bibnamefont {Asahi}},
  \bibinfo {author} {\bibfnamefont {A.}~\bibnamefont {Nagoya}}, \ and\ \bibinfo
  {author} {\bibfnamefont {G.}~\bibnamefont {Kresse}},\ }\href@noop {}
  {\bibfield  {journal} {\bibinfo  {journal} {Phys. Rev. B}\ }\textbf {\bibinfo
  {volume} {79}},\ \bibinfo {pages} {115126} (\bibinfo {year}
  {2009})}\BibitemShut {NoStop}%
\bibitem [{\citenamefont {Zhao}\ and\ \citenamefont {Persson}(2011)}]{123}%
  \BibitemOpen
  \bibfield  {author} {\bibinfo {author} {\bibfnamefont {H.}~\bibnamefont
  {Zhao}}\ and\ \bibinfo {author} {\bibfnamefont {C.}~\bibnamefont {Persson}},\
  }\href@noop {} {\bibfield  {journal} {\bibinfo  {journal} {Thin Solid Films}\
  }\textbf {\bibinfo {volume} {519}},\ \bibinfo {pages} {7508} (\bibinfo {year}
  {2011})}\BibitemShut {NoStop}%
\bibitem [{\citenamefont {Shen}\ \emph {et~al.}(2012)\citenamefont {Shen},
  \citenamefont {Jiang}, \citenamefont {Wang}, \citenamefont {Fu},
  \citenamefont {Zhou},\ and\ \citenamefont {Li}}]{Cu2ZnSnTe4}%
  \BibitemOpen
  \bibfield  {author} {\bibinfo {author} {\bibfnamefont {H.}~\bibnamefont
  {Shen}}, \bibinfo {author} {\bibfnamefont {X.~D.}\ \bibnamefont {Jiang}},
  \bibinfo {author} {\bibfnamefont {S.}~\bibnamefont {Wang}}, \bibinfo {author}
  {\bibfnamefont {Y.}~\bibnamefont {Fu}}, \bibinfo {author} {\bibfnamefont
  {C.}~\bibnamefont {Zhou}}, \ and\ \bibinfo {author} {\bibfnamefont {L.~S.}\
  \bibnamefont {Li}},\ }\href@noop {} {\bibfield  {journal} {\bibinfo
  {journal} {J. Mater. Chem.}\ }\textbf {\bibinfo {volume} {22}},\ \bibinfo
  {pages} {25050} (\bibinfo {year} {2012})}\BibitemShut {NoStop}%
\bibitem [{\citenamefont {Liu}\ \emph {et~al.}(2009{\natexlab{b}})\citenamefont
  {Liu}, \citenamefont {Huang}, \citenamefont {Chen}, ,\ and\ \citenamefont
  {Chen}}]{120}%
  \BibitemOpen
  \bibfield  {author} {\bibinfo {author} {\bibfnamefont {M.~L.}\ \bibnamefont
  {Liu}}, \bibinfo {author} {\bibfnamefont {F.~Q.}\ \bibnamefont {Huang}},
  \bibinfo {author} {\bibfnamefont {L.~D.}\ \bibnamefont {Chen}}, , \ and\
  \bibinfo {author} {\bibfnamefont {I.}~\bibnamefont {Chen}},\ }\href@noop {}
  {\bibfield  {journal} {\bibinfo  {journal} {Appl. Phys. Lett.}\ }\textbf
  {\bibinfo {volume} {94}},\ \bibinfo {pages} {202103} (\bibinfo {year}
  {2009}{\natexlab{b}})}\BibitemShut {NoStop}%
\bibitem [{\citenamefont {Jambor}\ and\ \citenamefont
  {Roberts}(1999)}]{Cu2CdGeS4}%
  \BibitemOpen
  \bibfield  {author} {\bibinfo {author} {\bibfnamefont {J.~L.}\ \bibnamefont
  {Jambor}}\ and\ \bibinfo {author} {\bibfnamefont {A.~C.}\ \bibnamefont
  {Roberts}},\ }\href@noop {} {\bibfield  {journal} {\bibinfo  {journal} {Am.
  Mineral.}\ }\textbf {\bibinfo {volume} {84}},\ \bibinfo {pages} {1464}
  (\bibinfo {year} {1999})}\BibitemShut {NoStop}%
\bibitem [{\citenamefont {Piskach}\ \emph {et~al.}(2000)\citenamefont
  {Piskach}, \citenamefont {Parasyuk},\ and\ \citenamefont {Romanyuk}}]{136}%
  \BibitemOpen
  \bibfield  {author} {\bibinfo {author} {\bibfnamefont {L.~V.}\ \bibnamefont
  {Piskach}}, \bibinfo {author} {\bibfnamefont {O.~V.}\ \bibnamefont
  {Parasyuk}}, \ and\ \bibinfo {author} {\bibfnamefont {Y.~E.}\ \bibnamefont
  {Romanyuk}},\ }\href@noop {} {\bibfield  {journal} {\bibinfo  {journal} {J.
  Allo. Comp.}\ }\textbf {\bibinfo {volume} {299}},\ \bibinfo {pages} {227}
  (\bibinfo {year} {2000})}\BibitemShut {NoStop}%
\bibitem [{\citenamefont {Schafer}\ and\ \citenamefont
  {Nitsche}(1977)}]{Cu2CdSnS4}%
  \BibitemOpen
  \bibfield  {author} {\bibinfo {author} {\bibfnamefont {W.}~\bibnamefont
  {Schafer}}\ and\ \bibinfo {author} {\bibfnamefont {R.}~\bibnamefont
  {Nitsche}},\ }\href@noop {} {\bibfield  {journal} {\bibinfo  {journal} {Z.
  Kristallogr.}\ }\textbf {\bibinfo {volume} {145}},\ \bibinfo {pages} {356}
  (\bibinfo {year} {1977})}\BibitemShut {NoStop}%
\bibitem [{\citenamefont {Guan}\ \emph {et~al.}(2013)\citenamefont {Guan},
  \citenamefont {Zhao}, \citenamefont {Wang},\ and\ \citenamefont {Yu}}]{140}%
  \BibitemOpen
  \bibfield  {author} {\bibinfo {author} {\bibfnamefont {H.}~\bibnamefont
  {Guan}}, \bibinfo {author} {\bibfnamefont {J.}~\bibnamefont {Zhao}}, \bibinfo
  {author} {\bibfnamefont {X.}~\bibnamefont {Wang}}, \ and\ \bibinfo {author}
  {\bibfnamefont {F.}~\bibnamefont {Yu}},\ }\href@noop {} {\bibfield  {journal}
  {\bibinfo  {journal} {Chalcogenide Lett.}\ }\textbf {\bibinfo {volume}
  {10}},\ \bibinfo {pages} {367} (\bibinfo {year} {2013})}\BibitemShut
  {NoStop}%
\bibitem [{\citenamefont {Hahn}\ and\ \citenamefont
  {Schulze}(1965)}]{stannite1}%
  \BibitemOpen
  \bibfield  {author} {\bibinfo {author} {\bibfnamefont {H.}~\bibnamefont
  {Hahn}}\ and\ \bibinfo {author} {\bibfnamefont {H.}~\bibnamefont {Schulze}},\
  }\href@noop {} {\bibfield  {journal} {\bibinfo  {journal}
  {Naturwissenschaften}\ }\textbf {\bibinfo {volume} {52}},\ \bibinfo {pages}
  {426} (\bibinfo {year} {1965})}\BibitemShut {NoStop}%
\bibitem [{\citenamefont {Ib\'{a}nez}\ \emph {et~al.}(2012)\citenamefont
  {Ib\'{a}nez}, \citenamefont {Cadavid}, \citenamefont {Zamani}, \citenamefont
  {Castell\'{o}}, \citenamefont {Roca}, \citenamefont {Li}, \citenamefont
  {Fairbrother}, \citenamefont {Prades}, \citenamefont {Shavel}, \citenamefont
  {Arbiol}, \citenamefont {Rodr\'{\i}guez}, \citenamefont {Morante},\ and\
  \citenamefont {Cabot}}]{144}%
  \BibitemOpen
  \bibfield  {author} {\bibinfo {author} {\bibfnamefont {M.}~\bibnamefont
  {Ib\'{a}nez}}, \bibinfo {author} {\bibfnamefont {D.}~\bibnamefont {Cadavid}},
  \bibinfo {author} {\bibfnamefont {R.}~\bibnamefont {Zamani}}, \bibinfo
  {author} {\bibfnamefont {N.~G.}\ \bibnamefont {Castell\'{o}}}, \bibinfo
  {author} {\bibfnamefont {V.~I.}\ \bibnamefont {Roca}}, \bibinfo {author}
  {\bibfnamefont {W.}~\bibnamefont {Li}}, \bibinfo {author} {\bibfnamefont
  {A.}~\bibnamefont {Fairbrother}}, \bibinfo {author} {\bibfnamefont {J.~D.}\
  \bibnamefont {Prades}}, \bibinfo {author} {\bibfnamefont {A.}~\bibnamefont
  {Shavel}}, \bibinfo {author} {\bibfnamefont {J.}~\bibnamefont {Arbiol}},
  \bibinfo {author} {\bibfnamefont {A.~P.}\ \bibnamefont {Rodr\'{\i}guez}},
  \bibinfo {author} {\bibfnamefont {J.~R.}\ \bibnamefont {Morante}}, \ and\
  \bibinfo {author} {\bibfnamefont {A.}~\bibnamefont {Cabot}},\ }\href@noop {}
  {\bibfield  {journal} {\bibinfo  {journal} {Chem. Mater.}\ }\textbf {\bibinfo
  {volume} {24}},\ \bibinfo {pages} {562} (\bibinfo {year} {2012})}\BibitemShut
  {NoStop}%
\bibitem [{\citenamefont {Haeuseler}\ \emph {et~al.}(1991)\citenamefont
  {Haeuseler}, \citenamefont {Ohrendorf},\ and\ \citenamefont
  {Himmrich}}]{Cu2CdSnTe4}%
  \BibitemOpen
  \bibfield  {author} {\bibinfo {author} {\bibfnamefont {H.}~\bibnamefont
  {Haeuseler}}, \bibinfo {author} {\bibfnamefont {F.~W.}\ \bibnamefont
  {Ohrendorf}}, \ and\ \bibinfo {author} {\bibfnamefont {M.}~\bibnamefont
  {Himmrich}},\ }\href@noop {} {\bibfield  {journal} {\bibinfo  {journal} {Z.
  Naturforsch. B}\ }\textbf {\bibinfo {volume} {46}},\ \bibinfo {pages} {1049}
  (\bibinfo {year} {1991})}\BibitemShut {NoStop}%
\bibitem [{\citenamefont {Sch\"{a}fer}\ and\ \citenamefont
  {Nitsche}(1974)}]{stannite2}%
  \BibitemOpen
  \bibfield  {author} {\bibinfo {author} {\bibfnamefont {W.}~\bibnamefont
  {Sch\"{a}fer}}\ and\ \bibinfo {author} {\bibfnamefont {R.}~\bibnamefont
  {Nitsche}},\ }\href@noop {} {\bibfield  {journal} {\bibinfo  {journal} {Mat.
  Res. Bull.}\ }\textbf {\bibinfo {volume} {9}},\ \bibinfo {pages} {645}
  (\bibinfo {year} {1974})}\BibitemShut {NoStop}%
\bibitem [{\citenamefont {Kabalov}\ \emph {et~al.}(1998)\citenamefont
  {Kabalov}, \citenamefont {Evstigneeva},\ and\ \citenamefont
  {Spiridonov}}]{Cu2HgSnS4}%
  \BibitemOpen
  \bibfield  {author} {\bibinfo {author} {\bibfnamefont {Y.~K.}\ \bibnamefont
  {Kabalov}}, \bibinfo {author} {\bibfnamefont {T.~L.}\ \bibnamefont
  {Evstigneeva}}, \ and\ \bibinfo {author} {\bibfnamefont {E.~M.}\ \bibnamefont
  {Spiridonov}},\ }\href@noop {} {\bibfield  {journal} {\bibinfo  {journal}
  {Krystallographiya}\ }\textbf {\bibinfo {volume} {43}},\ \bibinfo {pages}
  {21} (\bibinfo {year} {1998})}\BibitemShut {NoStop}%
\bibitem [{\citenamefont {Mkrtchyan}\ \emph {et~al.}(1988)\citenamefont
  {Mkrtchyan}, \citenamefont {Dovletov}, \citenamefont {Zhukov}, \citenamefont
  {Melikdzhanyan},\ and\ \citenamefont {Nuriev}}]{149_1}%
  \BibitemOpen
  \bibfield  {author} {\bibinfo {author} {\bibfnamefont {S.~A.}\ \bibnamefont
  {Mkrtchyan}}, \bibinfo {author} {\bibfnamefont {K.~O.}\ \bibnamefont
  {Dovletov}}, \bibinfo {author} {\bibfnamefont {E.~G.}\ \bibnamefont
  {Zhukov}}, \bibinfo {author} {\bibfnamefont {A.~G.}\ \bibnamefont
  {Melikdzhanyan}}, \ and\ \bibinfo {author} {\bibfnamefont {S.}~\bibnamefont
  {Nuriev}},\ }\href@noop {} {\bibfield  {journal} {\bibinfo  {journal} {Neorg.
  Mater.}\ }\textbf {\bibinfo {volume} {24}},\ \bibinfo {pages} {1094}
  (\bibinfo {year} {1988})}\BibitemShut {NoStop}%
\bibitem [{\citenamefont {Li}\ \emph {et~al.}(2012)\citenamefont {Li},
  \citenamefont {Zhang}, \citenamefont {Zhu}, \citenamefont {Zhang},\ and\
  \citenamefont {Ling}}]{146}%
  \BibitemOpen
  \bibfield  {author} {\bibinfo {author} {\bibfnamefont {D.}~\bibnamefont
  {Li}}, \bibinfo {author} {\bibfnamefont {X.}~\bibnamefont {Zhang}}, \bibinfo
  {author} {\bibfnamefont {Z.}~\bibnamefont {Zhu}}, \bibinfo {author}
  {\bibfnamefont {H.}~\bibnamefont {Zhang}}, \ and\ \bibinfo {author}
  {\bibfnamefont {F.}~\bibnamefont {Ling}},\ }\href@noop {} {\bibfield
  {journal} {\bibinfo  {journal} {Solid State Sci.}\ }\textbf {\bibinfo
  {volume} {14}},\ \bibinfo {pages} {890} (\bibinfo {year} {2012})}\BibitemShut
  {NoStop}%
\end{thebibliography}%

\end{document}